\documentclass[12pt,notitlepage]{amsart}%
\usepackage{amssymb}
\usepackage{amsfonts}
\usepackage{graphicx}
\usepackage{amscd}
\usepackage{graphicx}
\usepackage{amsmath}
\usepackage{amsxtra}
\usepackage{footmisc}%
\theoremstyle{plain}

\newtheorem{remark}{Remark}

\numberwithin{equation}{section}
\numberwithin{theorem}{section}
\numberwithin{lemma}{section}
\numberwithin{proposition}{section}
\numberwithin{corollary}{section}

\ifx\pdfoutput\relax\let\pdfoutput=\undefined\fi
\newcount\msipdfoutput
\ifx\pdfoutput\undefined\else
\ifcase\pdfoutput\else
\ifx\paperwidth\undefined\else
\ifdim\paperheight=0pt\relax\else\pdfpageheight\paperheight\fi
\ifdim\paperwidth=0pt\relax\else\pdfpagewidth\paperwidth\fi
\fi\fi\fi
\begin{document}
\title[QM is a non-local theory]{Quantum mechanics, non-locality, and the space discreteness hypothesis}
\author[Z\'{u}\~{n}iga-Galindo]{W. A. Z\'{u}\~{n}iga-Galindo}
\address{University of Texas Rio Grande Valley\\
School of Mathematical \& Statistical Sciences\\
One West University Blvd\\
Brownsville, TX 78520, United States}
\email{wilson.zunigagalindo@utrgv.edu}

\begin{abstract}
The space discreteness hypothesis asserts that the nature of space at short
distances is radically different from that at large distances. Based on the
Bronstein inequality, here, we use a totally disconnected topological space
$\mathcal{X}$ as a model for the physical space at short distances. However,
we consider the time as a real variable. In this framework, the Dirac-von
Neumann formalism can be used. This discreteness hypothesis implies that given
two different points in space, there is no continuous curve (a world line)
joining them. Consequently, this hypothesis is not compatible with the theory
of relativity. We propose $\mathbb{R}\times\left(  \mathbb{R}\times
\mathcal{X}\right)  ^{3}$ as a model of a space-time. For simplicity, we work
out our models with $\mathbb{R}\times\left(  \mathbb{R}\times\mathcal{X}%
\right)  $ as the configuration space. Quantum mechanics (QM), in the sense of
Dirac-von Neumann, on the Hilbert space $L^{2}\left(  \mathbb{R}%
\times\mathcal{X}\right)  $ is a non-local theory: the Hamiltonians are
non-local operators, and thus, spooky action at a distance is allowed. The
paradigm asserting that the universe is non-locally real implies that the
proposed version of QM admits realism. This version of QM can be specialized
to standard QM by using Hamiltonians acting on wavefunctions supported on the
region $\mathbb{R\times R}$. We apply the developed formalism to the
measurement problem. We propose a new mechanism for wavefunction collapse. The
mechanism resembles that proposed by Ghirardi, Rimini, and Weber, but there
are significant differences. The most important feature is that the
Schr\"{o}dinger equation describes the dynamics at all times, even at the
moment of measurement. We also discuss a model for the two-slit experiment in
which bright and dark states of light (recently proposed) naturally occur.

\end{abstract}
\maketitle

\section{Introduction}

Quantum mechanics (QM) is highly successful in describing the behavior of
matter and energy at the atomic and subatomic levels, which has driven the
development of technologies that have reshaped human society. However, there
is a consensus that QM has significant flaws; if the theory is not wrong, it
is at least incomplete. One major issue is the measurement problem,
specifically the collapse of the wave function during measurement. Other
issues are the incompatibility with general relativity and the existence of
several interpretations of quantum mechanics.

Non-locality of QM means that the components of a quantum system may continue
influence each other, even where they are well-separated and out of
speed-of-light contact. It was first pointed out by Einstein, Podolsky and
Rosen (EPR) in 1935, \cite{EPR}. EPR\ pointed out that non-locality implies
that the standard formalism of QM\ is incomplete or wrong. In the beginning of
the 1970s, Freedman and Clauser conducted experiments to test \ the
Bell-inequality violation, and other aspects of entangled quantum systems.
These experiments showed that both QM and the underlying reality it describes
are intrinsically non-local. Einstein's spooky actions-at-a-distance really
happened in the physical world. For an in-depth discussion, the reader may
consult \cite{Bell-1}-\cite{Grib and Rodriges}, and the references therein.

In the theory of relativity (TR), the use of manifolds that locally look like
$\mathbb{R}\times\mathbb{R}^{3}$ to describe the universe at large scales has
been highly successful. We argue that the clash between QM and the TR emerges
when the model of space-time $\mathbb{R}\times\mathbb{R}^{3}$ is used to
understand quantum phenomena. At the simplest possible level, causality
emerges from the total order of $\mathbb{R}$, and locality from the fact that
the standard derivatives define local operators, which in turn do not allow
spooky action at a distance.

The problem described above is know since 1920's; the creators of the QM were
convinced that the classical space-time continuum must abandoned on the
elementary quantum scale. Some them believed that the traditional concept of
space-time of macroscopic physics cannot simply be transferred to quantum
physics. But all the efforts to construct physical theories with discrete
space-times failed, \cite{Capellmann}.

In the 1930s, Bronstein showed that general relativity and quantum mechanics
imply that the uncertainty $\Delta x$ of any length measurement is greater
than or equal to the Planck length, \cite{Bronstein}. A widely accepted
interpretation of this inequality is that there are no intervals, just points
below the Planck length. We argue that this last assertion can be reformulated
in precise mathematical terms as follows: at the Planck length, the physical
space can be modeled as a totally disconnected space $\mathcal{X}$. In this
framework, there is no continuous world line joining two different points in
space $\mathcal{X}$; consequently, the assumption that $\mathcal{X}$ is a
model of space is entirely incompatible with the theory of relativity. In this
paper, \ $\mathcal{X}$ is the mathematical realization of the hypothesis of
the discreteness of space at small distances. The naive idea of using a
lattice in $\mathbb{R}^{3}$ as a model for the space discreteness is not
useful at all.

We propose $\mathbb{R}\times\left(  \mathbb{R}\times\mathcal{X}\right)  ^{3}$
as a model of the space-time (or more generally, $\mathbb{R}\times\left(
\mathbb{R}\times\mathcal{X}\right)  ^{N}$ as a model for the configuration
space). The advantage of this model is that it contains regions homeomorphic
to $\mathbb{R}\times\mathbb{R}^{3}$ and to $\mathbb{R}\times\mathcal{X}^{3}$.
In the first type of regions, causal and local theories like the relativistic
QM and TR can be formulated. In contrast, in the second type of regions,
acausal and non-local theories can be formulated. In this framework, \ QM (in
sense of Dirac-von Neumann) on the Hilbert spaces $L^{2}\left(  \left(
\mathbb{R}\times\mathcal{X}\right)  ^{N}\right)  $ can be rigorously
formulated, \cite{Dirac}-\cite{Takhtajan}. A rigorous mathematical study of QM
on Hilbert spaces of type $L^{2}\left(  \left(  \mathbb{R}\times
\mathcal{X}\right)  ^{N}\right)  $ is possible assuming suitable technical
conditions, but this choice is quite involved. Here, we take $N=1$,
$\mathcal{X}=\mathbb{Q}_{p}$, the field of $p$-adic numbers; $\mathbb{Q}_{p}$
is a paramount example of a totally disconnected space, with a very rich
mathematical structure. In the following paragraphs, we provide an argument
showing the physical relevance of the chosen model.

In the Dirac-von Neumann formulation of QM, the states of a closed quantum
system are vectors of an abstract complex Hilbert space $\mathcal{H}$, and the
observables correspond to linear self-adjoint operators in $\mathcal{H}$. In
$p$-adic QM, $\mathcal{H}=L^{2}(\mathbb{Q}_{p}^{N})$ or $\mathcal{H}=L^{2}%
(U)$, where $U\subset\mathbb{Q}_{p}^{N}$. This choice implies that the
position vectors are $\mathbb{Q}_{p}^{N}$, while the time is a real variable.
Using $\mathbb{R}\times\mathbb{Q}_{p}^{3}$ as a model of space-time, implies
assuming that the space ($\mathbb{Q}_{p}^{3}$)\ has a discrete nature: there
are no continuous word lines joining two different points in the space. We
warn the reader that there are several different types of $p$-adic QM; for
instance, if the time is assumed to be a $p$-adic variable, the QM obtained
radically differs from the one considered here. To the best of our knowledge,
$p$-adic QM (with $p$-adic time) started in the 1980s under the influence of
Vladimirov and Volovich, \cite{V-V-QM3}. The literature about $p$-adic QM is
pervasive, and here we cited just a few works, \cite{Volovich}%
-\cite{Zuniga-Chacon}.

A fundamental problem of $p$-adic quantum mechanics was the lack of physical
interpretation. In \cite{Zuniga-QM-2}, see also \cite{Zuniga-Mayes}%
-\cite{Zuniga-Chacon}, the author showed that a large class of $p$-adic
Schr\"{o}dinger equations is the scaling limit of certain continuous-time
quantum Markov chains (CTQMCs). This construction includes, as a particular
case, the continuous time quantum walks (CTQWs) constructed using adjacency
matrices, \cite{Childs et al}-\cite{Venegas-Andraca}. The CTQWs are a
fundamental tool in the development of quantum algorithms. The connection
between $p$-adic QM and CTQWs shows that $p$-adic QM has a physical meaning.
$p$-Adic QM is a non-local theory (the Hamiltonians used are non-local
operators), and consequently, spooky actions at a distance are allowed. The
paradigm asserting that the universe in not locally real, implies that
$p$-adic QM allows realism.

The non-locality tests are formulated using QM on $\mathcal{H}=\mathbb{C}^{N}%
$, see, e.g., \cite{several-authors}, \cite{Hardy}- \cite{Norsen}, and the
references therein.\ In Section \ref{Section_3}, we construct an
$N$-dimensional Hilbert space $\mathcal{\chi}_{N}(\mathbb{Z}_{p})$, spanned by
a set of functions whose supports are contained in the unit ball
$\mathbb{Z}_{p}$. Since $\mathcal{\chi}_{N}(\mathbb{Z}_{p})$ is isomorphic to
$\mathbb{C}^{N}$ as Hilbert spaces, the non-locality tests can be recast as
assertions in $p$-adic QM on the Hilbert space $L^{2}(\mathcal{\chi}%
_{N}(\mathbb{Z}_{p}))\subset L^{2}(\mathbb{Z}_{p})$. In addition, any
Hamiltonian on $\mathbb{C}^{N}\simeq\mathcal{\chi}_{N}(\mathbb{Z}_{p})$ (a
Hermitian matrix) has an extension as a non-local operator on $L^{2}%
(\mathbb{Z}_{p})$. In our view, this argument, and the connection with quantum
computing, establish the relevance of the choice $\mathcal{X}=\mathbb{Q}_{p}$.

Our first result is a theoretical framework, where QM is a non-local theory
admitting realism. An important step is to pass from a phenomenological
description of non-locality, as the one given in the second paragraph above,
to a mathematical one. We propose understanding non-locality in QM as the
emergence of the use of non-local operators (Hamiltonians) in quantum models.
These types of operators naturally allow spooky action at a distance, and then
this approach is consistent with the phenomenological one. However, the use of
local Hamiltonians (which do not enable spooky action at a distance)
contradicts the experimental confirmation of non-locality.

In Section \ref{Section_4}, we discuss the Schr\"{o}dinger equations that
control the dynamics of quantum states $\Psi\left(  x_{\infty},x_{p},t\right)
$, $x_{\infty}\in\mathbb{R}$, $x_{p}\in\mathbb{Q}_{p}$, $t\in\mathbb{R}$, with
$\Psi\left(  \cdot,\cdot,t\right)  \in L^{2}\left(  \mathbb{R\times Q}%
_{p}\right)  $. We assume that the standard Schr\"{o}dinger equation is valid
in the region $\mathbb{R\times R}$; which means that if $\Psi_{\infty}\left(
x_{\infty},t\right)  $, $x_{\infty}\in\mathbb{R}$, $t\in\mathbb{R}$, \ is a
quantum state from $L^{2}\left(  \mathbb{R}\right)  $, it evolves according to%
\[
i\frac{\partial}{\partial t}\Psi_{\infty}\left(  x_{\infty},t\right)
=\boldsymbol{H}_{\infty}\Psi_{\infty}\left(  x_{\infty},t\right)  ,
\]
where $\boldsymbol{H}_{\infty}:Dom(\boldsymbol{H}_{\infty})\rightarrow
L^{2}\left(  \mathbb{R}\right)  $ is a linear, local Hamiltonian densely
defined in $L^{2}\left(  \mathbb{R}\right)  $. We also assume that $p$-adic QM
is valid in the region $\mathbb{R\times Q}_{p}$; which means that if $\Psi
_{p}\left(  x_{p},t\right)  $, $x_{p}\in\mathbb{Q}_{p}$, $t\in\mathbb{R}$,
\ is a quantum state from $L^{2}\left(  \mathbb{Q}_{p}\right)  $, it evolves
according to%
\[
i\frac{\partial}{\partial t}\Psi_{p}\left(  x_{p},t\right)  =\boldsymbol{H}%
_{p}\Psi_{p}\left(  x_{p},t\right)  ,
\]
where $\boldsymbol{H}_{p}:Dom(\boldsymbol{H}_{p})\rightarrow L^{2}\left(
\mathbb{Q}_{p}\right)  $ is a linear, non-local Hamiltonian densely defined in
$L^{2}\left(  \mathbb{Q}_{p}\right)  $. Now, by using the ansatz%
\[
\Psi\left(  x_{\infty},x_{p},t\right)  =\Psi_{\infty}\left(  x_{\infty
},t\right)  \Psi_{p}\left(  x_{p},t\right)  ,
\]
which is motivated by the fact that $L^{2}\left(  \mathbb{R}\right)
{\textstyle\bigotimes}
L^{2}\left(  \mathbb{Q}_{p}\right)  \simeq L^{2}\left(  \mathbb{R\times Q}%
_{p}\right)  $, one finds easily that%
\begin{equation}
i\frac{\partial}{\partial t}\Psi\left(  x_{\infty},x_{p},t\right)  =\left(
\boldsymbol{H}_{p}+\boldsymbol{H}_{\infty}\right)  \Psi\left(  x_{\infty
},x_{p},t\right)  , \label{Large_Schodinger}%
\end{equation}
which implies that QM on $L^{2}\left(  \mathbb{R\times Q}_{p}\right)  $ is a
non-local theory, since $\boldsymbol{H}_{p}+\boldsymbol{H}_{\infty}$ is a
non-local operator. Then, the paradigm that the universe is non-locally real
implies that the proposed version of QM admits realism. Here, it is relevant
to mention that the above argument does not give all the admissible
Hamiltonians; see Section \ref{Section_4} for further details.

In Section \ref{Section_6}, we apply the theoretical framework developed to
the measurement problem. In this framework, the collapse of the wavefunction
is a consequence of the geometry of the space-time. The collapse mechanism
resembles the one proposed by Ghirardi, Ramini, and Weber (GRW); see, e.g.,
\cite{Norsen}-\cite{Bell -Againts-Meas}, but, there are three major
differences. First, the wavefunction is not a physical entity; second, we do
not need to introduce new physical constants. Third, the Schr\"{o}dinger
equation (\ref{Large_Schodinger}) described the collapse of the wavefunction,
so, it is not necessary to introduce a new dynamical equation.

In Section \ref{Section_7}, we discuss the two-slit experiment. This Section
is based on \cite{Zuniga-AP}, \cite{Webb}, and \cite{Villas-Boas et al}. Our
model gives two different interference patterns $\left\vert \Psi_{p}\left(
x_{p},t\right)  \right\vert ^{2}$, $\left\vert \Psi_{\infty}\left(  x_{\infty
},t\right)  \right\vert ^{2}$. The second one is in the macroscopic realm
($\mathbb{R\times R}$),\ while the first one is in the microscopic realm
($\mathbb{R\times Q}_{p}$). By interpreting $\Psi_{p}\left(  x_{p},t\right)  $
as dark state, and $\Psi_{\infty}\left(  x_{\infty},t\right)  $ as bright
state, we describe the two-slit experiment similar to the one given recently
in \cite{Villas-Boas et al}.

The model introduced here opens many questions and controversies. In Section
\ref{Section_8}, we discuss briefly some matters like Einstein's causality,
the no-communication theorem, and the nature of space-time at small distances, etc.

\section{The space discreteness hypothesis}

\subsection{The Bronstein inequality}

Since the 1920s, Born, Einstein, Jordan, Pauli, among others, were convinced
that the classical space-time continuum must abandoned on the elementary
quantum scale. The reader may consult \cite{Capellmann} for an in-depth
historical review. For instance, according to \cite{Capellmann}, Einstein and
Born believed that the traditional concept of space-time of macroscopic
physics cannot simply be transferred to quantum physics. But all the efforts
(especially from Born and Jordan) to construct physical theories with discrete
space-times failed.

In the 1930s, Bronstein showed that general relativity and quantum mechanics
imply that the uncertainty $\Delta x$ of any length measurement satisfies
\begin{equation}
\Delta x\geq L_{\text{Planck}}:=\sqrt{\frac{\hbar G}{c^{3}}},
\label{Inequality}%
\end{equation}
where $L_{\text{Planck}}$ is the Planck length ($L_{\text{Planck}}%
\approx10^{-33}$ $cm$), \cite{Bronstein}. The standard interpretation of
Bronstein's inequality is that below the Planck length, there are no
intervals, just points. The intervals are the basic examples of connected
sets; such a set is the mathematical incarnation of an infinitely divisible
object. Then, below the Planck length, the topology of space only allows
trivial (points) connected subsets. This can be recast as below the Planck
length, the space is a totally (or completely) disconnected topological space.

We denote by $\mathcal{X}$ a totally disconnected space. Notice $\mathbb{R}%
^{N}$ is connected space. Consider a word line (a continuous curve) in
$\mathcal{X}$:%
\[%
\begin{array}
[c]{cccc}%
x: & \left[  a,b\right]  & \rightarrow & \mathcal{X}\\
& t & \rightarrow & x\left(  t\right)  .
\end{array}
\]
We recall that continuous mappings preserve connectedness. Then, the image
$x\left(  \left[  a,b\right]  \right)  $ must be a connected subset of
$\mathcal{X}$, since the only connected subsets are points; the curve
$x\left(  t\right)  $ is constant. In other words, given two different points
in $\mathcal{X}$, there is no word line joining them. The motion in
$\mathcal{X}$ is a sequence of jumps. Consequently, we replace the space-time
model $\mathbb{R}\times\mathbb{R}^{3}$ by $\mathbb{R}\times\mathcal{X}^{3}$,
i.e., if we assume the discreteness of space, then we lose the compatibility
with special and general relativity.

\subsection{Macroscopic versus microscopic}

We divide the space-time into two realms: the macroscopic and the microscopic.
In the first one, the model of the space-time is $\mathbb{R}\times
\mathbb{R}^{3}$, while in the second one, it is $\mathbb{R}\times
\mathcal{X}^{3}$, where $\mathcal{X}$ is a totally disconnected topological
space. The connectedness of a topological space is a topological property,
which means it remains invariant under continuous transformations
(homeomorphisms). Therefore, $\mathcal{X}$ is not topologically equivalent to
$\mathbb{R}$, nor a proper subset of $\mathbb{R}$.

The formulation of models using $\mathbb{R}\times\mathcal{X}^{N}$ as a
configuration space requires that algebraic operations, like sum and product,
be defined on $\mathcal{X}$. A natural candidate for $\mathcal{X}$ is a
$K$-analytic Lie group, where $K$ is a non-Archimedean field, \cite{Serre}. It
is not possible to assume a sharp boundary between the macroscopic and
microscopic realms, and, in addition, there is no topologically faithful copy
of $\mathcal{X}$ in $\mathbb{R}$; for this reason, we propose
\[
\mathbb{R}\times\left(  \mathbb{R}\times\mathcal{X}\right)  ^{3}%
\]
as a model of space-time. Due to mathematical convenience, we use the word
space-time in an extended way to mean configuration space. Then, we propose
using configuration spaces of type $\mathbb{R}\times\left(  \mathbb{R}%
\times\mathcal{X}\right)  ^{N}$.

\subsection{The $p$-adic space}

In this section, we fix the notation and collect some basic results on
$p$-adic analysis that we use in this paper. For a detailed exposition on
$p$-adic analysis, the reader may consult \cite{V-V-Z}, \cite{Zuniga-Textbook}%
-\cite{Alberio et al}. For a discussion on applications of $p$-adic analysis
to physics, biology, computing, and other areas, the reader may consult
\cite{Anashin}, \cite{Zuniga-AP}, \cite{Zuniga-Chacon}, \cite{R-T-V}%
-\cite{Zuniga-networks}, \cite{Zuniga-Galindo-JNMP}, and the references therein.

From now on, we use $p$ to denote a fixed prime number. Any non-zero $p$-adic
number $x$ has a unique expansion of the form%
\begin{equation}
x=x_{-k}p^{-k}+x_{-k+1}p^{-k+1}+\ldots+x_{0}+x_{1}p+\ldots,\text{ }
\label{p-adic-number}%
\end{equation}
with $x_{-k}\neq0$, where $k$ is an integer, and the $x_{j}$s\ are numbers
from the set $\left\{  0,1,\ldots,p-1\right\}  $. The set of all possible
sequences of the form (\ref{p-adic-number}) constitutes the field of $p$-adic
numbers $\mathbb{Q}_{p}$. There are natural field operations, sum and
multiplication, on series of form (\ref{p-adic-number}). There is also a norm
in $\mathbb{Q}_{p}$ defined as $\left\vert x\right\vert _{p}=p^{-ord(x)}$,
where $ord_{p}(x)=ord(x)=k$, for a nonzero $p$-adic number $x$. By definition
$ord(0)=\infty$. The field of $p$-adic numbers with the distance induced by
$\left\vert \cdot\right\vert _{p}$ is a complete ultrametric space. The
ultrametric property refers to the fact that $\left\vert x-y\right\vert
_{p}\leq\max\left\{  \left\vert x-z\right\vert _{p},\left\vert z-y\right\vert
_{p}\right\}  $ for any $x$, $y$, $z$ in $\mathbb{Q}_{p}$. The $p$-adic
integers which are sequences of form (\ref{p-adic-number}) with $-k\geq0$. All
these sequences constitute the unit ball $\mathbb{Z}_{p}$. The unit ball is an
infinite rooted tree with fractal structure. As a topological space
$\mathbb{Q}_{p}$\ is homeomorphic to a Cantor-like subset of the real line,
see, e.g., \cite{V-V-Z}, \cite{Alberio et al}. $\mathbb{Q}_{p}$ is a paramount
example of a totally disconnected space, with a very rich mathematical structure.

A function $\varphi:\mathbb{Q}_{p}\rightarrow\mathbb{C}$ is called locally
constant, if for any $a\in\mathbb{Q}_{p}$, there is an integer $l=l(a)$, such
that
\[
\varphi\left(  a+x\right)  =\varphi\left(  a\right)  \text{ for any }%
|x|_{p}\leq p^{l}.
\]
The set of functions for which $l=l\left(  \varphi\right)  $ depends only on
$\varphi$ form a $\mathbb{C}$-vector space denoted as $\mathcal{U}%
_{loc}\left(  \mathbb{Q}_{p}\right)  $. We call $l\left(  \varphi\right)  $
the exponent of local constancy. If $\varphi\in\mathcal{U}_{loc}\left(
\mathbb{Q}_{p}\right)  $ has compact support, we say that $\varphi$ is a test
function. We denote by $\mathcal{D}(\mathbb{Q}_{p})$ the complex vector space
of test functions. There is a natural integration theory so that
$\int_{\mathbb{Q}_{p}}\varphi\left(  x\right)  dx$ gives a well-defined
complex number. The measure $dx$ is the Haar measure of $\mathbb{Q}_{p}$.

Let $U\subset\mathbb{Q}_{p}$ be an open subset. We denote by $\mathcal{D}(U)$
denotes the $\mathbb{C}$-vector space of test functions with supports
contained in $U$; then, $\mathcal{D}(U)$ is dense in
\[
L^{\rho}(U)=\left\{  \varphi:(U)\rightarrow\mathbb{C};\left\Vert
\varphi\right\Vert _{\rho}=\left\{
{\displaystyle\int\limits_{U}}
\left\vert \varphi\left(  x\right)  \right\vert ^{\rho}dx\right\}  ^{\frac
{1}{\rho}}<\infty\right\}  ,
\]
for $1\leq\rho<\infty$, see, e.g., \cite[Section 4.3]{Alberio et al}.

We now take $U=\mathbb{Z}_{p}$ and $\rho=2$. The density $\mathcal{D}%
(\mathbb{Z}_{p})$ in $L^{2}\left(  \mathbb{Z}_{p}\right)  $ means that given
$f\in L^{2}\left(  \mathbb{Z}_{p}\right)  $ and $\epsilon>0$, there exist a
nonnegative integer $l$, and
\[
\varphi^{\left(  l\right)  }\left(  x\right)  =\sum_{I\in G_{l}}\varphi
_{I,l}\Omega\left(  p^{l}\left\vert x-I\right\vert _{p}\right)  \in
\mathcal{D}(\mathbb{Z}_{p}),
\]
such that $\left\Vert f-\varphi^{\left(  l\right)  }\right\Vert _{2}<\epsilon
$. Here, $I=I_{0}+\ldots+I_{l-1}p^{l-1}$, with $I_{k}\in\left\{
0,\ldots,p-1\right\}  $, $\varphi_{I,l}\in\mathbb{C}$, and $\Omega\left(
p^{l}\left\vert x-I\right\vert _{p}\right)  $ is the characteristic function
of the ball $I+p^{l}\mathbb{Z}_{p}$. The function $\varphi^{\left(  l\right)
}$ is a discretization of $f$; the functions $\varphi^{\left(  l\right)  }$
constitute a $\mathbb{C}$-vector space $\mathcal{D}_{l}(\mathbb{Z}_{p})$,
which is isometric to the finite-dimensional Hilbert space $\mathbb{C}^{p^{l}%
}$. The functions $\left\{  p^{\frac{l}{2}}\Omega\left(  p^{l}\left\vert
x-I\right\vert _{p}\right)  \right\}  _{I\in G_{l}}$ form an orthonormal
basis, indeed%
\begin{gather*}
\left\langle p^{\frac{l}{2}}\Omega\left(  p^{l}\left\vert x-I\right\vert
_{p}\right)  ,p^{\frac{l}{2}}\Omega\left(  p^{l}\left\vert x-J\right\vert
_{p}\right)  \right\rangle =p^{l}%
{\displaystyle\int\limits_{\mathbb{Z}_{p}}}
\Omega\left(  p^{l}\left\vert x-I\right\vert _{p}\right)  \overline
{\Omega\left(  p^{l}\left\vert x-J\right\vert _{p}\right)  }dx\\
=p^{l}%
{\displaystyle\int\limits_{\left(  I+p^{l}\mathbb{Z}_{p}\right)  \cap\left(
J+p^{l}\mathbb{Z}_{p}\right)  }}
dx=p^{l}\delta_{I,J}%
{\displaystyle\int\limits_{I+p^{l}\mathbb{Z}_{p}}}
dx=\delta_{I,J},
\end{gather*}
where $\delta_{I,J}$ denotes the Kronecker delta.

\subsection{$p$-Adic QM}

In the Dirac-Von Neumann formulation of QM, \cite{Berezin et al}%
-\cite{Takhtajan}, to every isolated quantum system there is associated a
separable complex Hilbert space $\mathcal{H}$ called the space of states. The
states of a quantum system are described by non-zero vectors from
$\mathcal{H}$. Each observable corresponds to a unique linear self-adjoint
operator in $\mathcal{H}$. The most important observable of a quantum system
is its energy. We denote the corresponding operator by $\boldsymbol{H}$. Let
$\Psi_{0}\in\mathcal{H}$ be the state at time $t=0$ of a certain quantum
system. Then at time $t$ the system is represented by the vector $\Psi\left(
t\right)  =\boldsymbol{U}_{t}\Psi_{0}$, where $\boldsymbol{U}_{t}%
=e^{-it\boldsymbol{H}}$, $t\geq0$,\ is a unitary operator called the evolution
operator. The vector function $\Psi\left(  t\right)  $ is differentiable if
$\Psi\left(  t\right)  $ is contained in the domain $Dom(\boldsymbol{H})$ of
$\boldsymbol{H}$, which happens\ if at $t=0$, $\Psi_{0}\in Dom(\boldsymbol{H}%
)$, and in this case the time evolution of $\Psi\left(  t\right)  $ is
controlled by the Schr\"{o}dinger equation $i\frac{\partial}{\partial t}%
\Psi\left(  t\right)  =\boldsymbol{H}\Psi\left(  t\right)  $, where
$i=\sqrt{-1}$ and the Planck constant is assumed to be one.

In $p$-adic QM, $\mathcal{H=}L^{2}\left(  U\right)  $, where $U$ is an open
compact subset of $\mathbb{Q}_{p}^{N}$. Notice that the spatial variables are
vectors in $\mathbb{Q}_{p}^{N}$, but the time is real variable. This last
\ condition is completely necessary to use the Dirac-von Neumann formalism.
Now, the Hilbert spaces $L^{2}\left(  \mathbb{Q}_{p}^{3}\right)  $ and
$L^{2}\left(  \mathbb{R}^{3}\right)  $ are isometric, because both have
countable orthonormal bases; but the QM on $\mathcal{H=}L^{2}\left(
\mathbb{R}^{3}\right)  $ is a relativistic theory, but QM on $\mathcal{H=}%
L^{2}\left(  \mathbb{Q}_{p}^{3}\right)  $ is not. Then, the geometry of the
configuration space changes radically the properties of the quantum models considered.

A natural `philosophical' question is the existence of a universal model
$\mathcal{X}$ for the space at the microscopic level. Here, we argue that in
QM, the selection of a Hilbert space also includes the selection of a
configuration space $\mathcal{X}$, and that the selection is made considering
the particular phenomena to be modeled. This is a well established practice in
QM, as Berezin and Shubin explain in their book \cite{Berezin et al}:
\textquotedblleft The question of how, for a concrete physical system, to
describe the state space $\mathcal{H}$, and establish a correspondence between
the observables and self-adjoint operators in $\mathcal{H}$ goes beyond a
purely mathematical theory and belongs to the domain of physical practice and
intuition. In every particular case this question should be handled by an
expert physicist.\textquotedblright

\subsection{Non-locality in $p$-adic QM}

As discussed in the introduction, here, it is more convenient to understand
non-locality as the emergence of non-local operators in physical models (the
Hamiltonians are non-local). Non-local Hamiltonians allow spooky action at a
distance. In the phenomenological approach, non-locality means the existence
of correlations between separated particles that are stronger than what
classical physics allows. This, in turn, implies that measuring a property of
one particle can instantaneously influence the properties of another, even if
they are far apart. This interpretation of non-locality is entirely consistent
with the hypothesis of non-local Hamiltonians; however, it comes into
contradiction with the hypothesis that the Hamiltonians are local operators.
Here, it is relevant to recall that the locality of the Lagrangian densities
is a required hypothesis in the formulation of functional integrals in field
theory, \cite[Section 5.1]{Zinn-Justin}.

Along this paper, we use discrete space as a synonymous of totally
disconnected space. The non-locality is natural in discrete spaces like
$\mathbb{Q}_{p}$. On a totally disconnected topological space, like
$\mathbb{Q}_{p}$, the locally constant functions play a central role. A
function $\varphi:\mathbb{Q}_{p}\rightarrow\mathbb{C}$ is locally constant, if
$\varphi\left(  x+x^{\prime}\right)  -\varphi\left(  x\right)  =0$ for
$x^{\prime}$ sufficiently small. Then, the standard notion of derivative
defined as%
\[
\lim_{x^{\prime}\rightarrow0}\frac{\varphi\left(  x+x^{\prime}\right)
-\varphi\left(  x\right)  }{\left\vert x^{\prime}\right\vert _{p}}=0
\]
is not useful. However, in the $p$-adic framework, the notion of \ fractional
derivative is available. The simplest Hamiltonian is $\boldsymbol{D}^{\alpha}%
$, $\alpha>0$, the Taibleson-Vladimirov fractional, which is defined as%
\begin{equation}
\boldsymbol{D}^{\alpha}\varphi\left(  x\right)  =\frac{1-p^{\alpha}%
}{1-p^{-\alpha-1}}%
{\displaystyle\int\limits_{\mathbb{Q}_{p}}}
\frac{\varphi(z)-\varphi(x)}{|z-x|_{p}^{\alpha+1}}\,dz,
\label{Vladimirov-Taibleson-Derivative}%
\end{equation}
for $\varphi$ a locally constant function with compact support; see, e.g.,
\cite[Chapter 2]{Zuniga-Textbook}. To see the non-local nature of this
operator, we take $\varphi\left(  x\right)  =1$ if $|x|_{p}\leq1$, otherwise
$\varphi\left(  x\right)  =0$, then%
\begin{align*}
\boldsymbol{D}^{\alpha}\varphi\left(  x\right)   &  =\frac{1-p^{\alpha}%
}{1-p^{-\alpha-1}}\left\{
{\displaystyle\int\limits_{|z|_{p}\leq1}}
\frac{\varphi(z)-\varphi(x)}{|z-x|_{p}^{\alpha+1}}\,dz+%
{\displaystyle\int\limits_{|z|_{p}>1}}
\frac{\varphi(z)-\varphi(x)}{|z-x|_{p}^{\alpha+1}}\,dz\right\} \\
&  =\left\{
\begin{array}
[c]{lll}%
-\frac{1-p^{\alpha}}{1-p^{-\alpha-1}}\left(  \text{ }%
{\displaystyle\int\limits_{|z|_{p}>1}}
\frac{dz}{|z|_{p}^{\alpha+1}}\,\right)  & \text{if} & |x|_{p}\leq1\\
&  & \\
\frac{1-p^{\alpha}}{1-p^{-\alpha-1}}\frac{1}{|x|_{p}^{\alpha+1}} & \text{if} &
|x|_{p}>1.
\end{array}
\right.
\end{align*}
Then, the Hamiltonian $\boldsymbol{D}^{\alpha}$ allows spooky action at a distance.

\subsection{Ultrametricity}

Ultrametricity in physics means the emergence of ultrametric spaces in
physical models. A metric space $(M,d)$ is ultrametric if \ the distance $d$
satisfies $d(A,B)\leq\max\left\{  d(A,C),d(C,B)\right\}  $ for any three
points $A$, $B,$ $C$ in $M$. Ultrametricity was discovered in the 80s by
Parisi and others in the theory of spin glasses and by Frauenfelder and others
in physics of proteins; see, e.g.,\ \cite{Fraunfelder et al}, \cite{R-T-V},
see also \cite[Chapter 4]{KKZuniga}, and \ the references therein. The
$p$-adic heat equations are continuous versions (scaling limits) of master
equations of complex hierarchical systems, \cite{Zuniga-networks},
\cite{Zuniga-LNM-2016}-\cite{zuniga2020reaction}. By performing a Wick
rotation, in a $p$-adic heat equation, one gets a $p$-adic Schr\"{o}dinger
equation. This type of equations can be interpreted as \ scaling limits of
continuous-time quantum Markov chains, \cite{Zuniga-QM-2},
\cite{Zuniga-Chacon}, and the references therein.

\section{\label{Section_3}$p$-Adic QM is the quantum model for quantum
computing}

In \cite{Zuniga-QM-2}, see also \cite{Zuniga-Chacon}, we showed that a large
class of $p$-adic Schr\"{o}dinger equations is the scaling limit of certain
continuous-time quantum Markov chains (CTQMCs). Practically, a discretization
of such an equation gives a CTQMC. As a practical result, we construct new
types of continuous-time quantum walks (CTQWs) on graphs using two symmetric
matrices. This construction includes, as a particular case, the CTQWs
constructed using adjacency matrices, \cite{Childs et al}-\cite{Mulkne-Blumen}%
, \cite{Venegas-Andraca}. The connection between $p$-adic QM and CTQWs show
that $p$-adic QM has a physical meaning. $p$-Adic QM is a non-local theory
because the Hamiltonians used are non-local operators, and consequently,
spooky actions at a distance are allowed. But, this theory is not a
mathematical toy. The paradigm asserting that the universe in not locally
real, implies that $p$-adic QM allows realism.

Quantum computing and the non-locality tests can be formulated using $p$-adic
QM. This idea is already presented\ in \cite{Zuniga-QM-2}. Here, we give a
quick review. The standard formulation of these theories uses QM on the
Hilbert space $\mathbb{C}^{N}$. We select a prime number $p$ and a positive
integer $l$, and a subset $G_{l}^{0}\subseteq G_{l}$, with cardinality
$\#G_{l}^{0}=N$. Notice that we require that $N\leq p^{l}$. Now, the set%
\begin{equation}
\left\{  p^{\frac{l}{2}}\Omega\left(  p^{l}\left\vert x-I\right\vert
_{p}\right)  ;I\in G_{l}^{0}\right\}  \subset\mathcal{D}_{l}(\mathbb{Z}_{p})
\label{basis}%
\end{equation}
is orthonormal set with respect to the inner product%
\[
\left\langle \varphi,\phi\right\rangle =%
{\displaystyle\int\limits_{\mathbb{Z}_{p}}}
\varphi\left(  x\right)  \overline{\phi}\left(  x\right)  dx.
\]
We denote by $\mathcal{\chi}_{N}(\mathbb{Z}_{p})$ the $N$-dimensional Hilbert
space spanned by (\ref{basis}). Then, $\mathcal{\chi}_{N}(\mathbb{Z}_{p})$ and
$\mathbb{C}^{N}$ are isomorphic as Hilbert spaces. Notice that the isomorphism
exists for any $p$ and $l$ such that $N\leq p^{l}$.

The advantage of using $\mathcal{\chi}_{N}(\mathbb{Z}_{p})$ over
$\mathbb{C}^{N}$ is that we can construct, in a natural way, continuous
versions of quantum models in $\mathbb{C}^{N}$. We identify $p^{\frac{l}{2}%
}\Omega\left(  p^{l}\left\vert x-I\right\vert _{p}\right)  $ with $e_{I}$,
where $\left\{  e_{I}\right\}  _{I\in G_{l}^{0}}$ is the canonical basis of
$\mathbb{C}^{N}$. Let
\[
\left[  H_{J,K}\right]  _{1\leq J,K\leq N}=\left[  H_{J,K}\right]  _{J,K\in
G_{l}^{0}}%
\]
be a Hermitian matrix. We attach to this matrix the kernel%
\[
h(x,y):=p^{l}%
{\displaystyle\sum\limits_{J\in G_{l}^{0}}}
{\displaystyle\sum\limits_{K\in G_{l}^{0}}}
H_{J,K}\Omega\left(  p^{l}\left\vert x-J\right\vert _{p}\right)  \Omega\left(
p^{l}\left\vert y-K\right\vert _{p}\right)  ,
\]
for $x,y\in\mathbb{Z}_{p}$, and the linear operator%
\begin{equation}
\varphi\left(  x\right)  \rightarrow\boldsymbol{H}\varphi\left(  x\right)  =%
{\displaystyle\int\limits_{\mathbb{Z}_{p}}}
h(x,y)\varphi\left(  y\right)  dy, \label{Integral}%
\end{equation}
where $\varphi\left(  y\right)  =\sum_{K\in G_{l}^{0}}c_{K}\Omega\left(
p^{l}\left\vert y-K\right\vert _{p}\right)  \in\mathcal{\chi}_{N}%
(\mathbb{Z}_{p})$. \ Now, if we take $\varphi\in\mathcal{C}(\mathbb{Z}_{p})$
to be a continuous function defined on $\mathbb{Z}_{p}$, then, the integral in
(\ref{Integral}) gives a function from $\mathcal{C}(\mathbb{Z}_{p})$.
Furthermore, $\mathcal{C}(\mathbb{Z}_{p})$ is dense in $L^{2}(\mathbb{Z}_{p}%
)$, and$\mathcal{\ }$
\[
\left\Vert \boldsymbol{H}\varphi\right\Vert _{L^{2}(\mathbb{Z}_{p})}%
\leq\left\Vert h\right\Vert _{L^{2}(\mathbb{Z}_{p}\times\mathbb{Z}_{p}%
)}\left\Vert \varphi\right\Vert _{L^{2}(\mathbb{Z}_{p})}.
\]
Therefore, $\boldsymbol{H}$ extends \ to a linear, bounded operator on
$L^{2}(\mathbb{Z}_{p})$. Notice that this operator is non-local, and that by
construction $\boldsymbol{H}:\mathcal{\chi}_{N}(\mathbb{Z}_{p})\rightarrow
\mathcal{\chi}_{N}(\mathbb{Z}_{p})$; this operator is represented by the
matrix $\left[  H_{J,K}\right]  _{1\leq J,K\leq N}$. It is not difficult to
verify that $\boldsymbol{H}$ is symmetric, and consequently self-adjoint. In
conclusion,%
\[
i\frac{\partial}{\partial t}\Psi\left(  x,t\right)  =\boldsymbol{H}\Psi\left(
x,t\right)  \text{, }x\in\mathbb{Z}_{p}\text{, }t\geq0,
\]
is a continuous Schr\"{o}dinger equation attached to matrix $\left[
H_{J,K}\right]  _{1\leq J,K\leq N}$. For further details, the reader may
consult \cite{Zuniga-QM-2}; see also \cite{Zuniga-Mayes}-\cite{Zuniga-Chacon},
and the references therein.

In 1964, Bell established that quantum mechanics is inherently non-local
through a non-locality test known now as Bell inequalities, \cite{Bell-1}.
Later, non-locality tests without using inequalities were established by
several authors; see, e.g., \cite{several-authors}, \cite{Hardy}. A relevant
observation is that QM on $L^{2}(\mathcal{\chi}_{N}(\mathbb{Z}_{p}))\subset
L^{2}(\mathbb{Z}_{p})$ is a non-local theory, which naturally agrees with
Bell's conclusion.

\section{\label{Section_4}QM is a non-local theory that admits realism}

From now on, we assume that $\mathcal{X}=\mathbb{Q}_{p}$. For the sake of the
simplicity, we work in the configuration space $\mathbb{R}\times\left(
\mathbb{R\times Q}_{p}\right)  $. There are two reasons for this choice.
First, the $p$-adic QM is a relevant formalism due to its connection with
quantum computing. Second, the discussion and results can be easily extended
to more general spaces by introducing suitable hypotheses.

\subsection{An orthonormal basis for $L^{2}(\mathbb{Q}_{p})$}

Any $p$-adic number $x\neq0$ has a unique expansion of the form $x=p^{ord(x)}%
\sum_{j=0}^{\infty}x_{j}p^{j},$ where $x_{j}\in\{0,\dots,p-1\}$ and $x_{0}%
\neq0$. By using this expansion, we define the fractional part of\textit{
}$x\in\mathbb{Q}_{p}$, denoted $\{x\}_{p}$, as the rational number
\[
\left\{  x\right\}  _{p}=\left\{
\begin{array}
[c]{lll}%
0 & \text{if} & x=0\text{ or }ord(x)\geq0\\
&  & \\
p^{ord(x)}\sum_{j=0}^{-ord(x)-1}x_{j}p^{j} & \text{if} & ord(x)<0.
\end{array}
\right.
\]
Set $\chi_{p}(y)=\exp(2\pi i\{y\}_{p})$ for $y\in\mathbb{Q}_{p}$. The map
$\chi_{p}(\cdot)$ is an additive character on $\mathbb{Q}_{p}$, i.e., a
continuous map from $\left(  \mathbb{Q}_{p},+\right)  $ into $S$ (the unit
circle considered as a multiplicative group) satisfying $\chi_{p}(x_{0}%
+x_{1})=\chi_{p}(x_{0})\chi_{p}(x_{1})$, $x_{0},x_{1}\in\mathbb{Q}_{p}$; see,
e.g., \cite[Section 2.3]{Alberio et al}.

We set%
\[
\mathbb{Q}_{p}/\mathbb{Z}_{p}=\left\{  \sum_{j=-1}^{-m}x_{j}p^{j};\text{for
some }m>0\right\}  .
\]
For $b\in\mathbb{Q}_{p}/\mathbb{Z}_{p}$, $r\in\mathbb{Z}$, we denote by
$\Omega\left(  \left\vert p^{r}x_{p}-b\right\vert _{p}\right)  $ the
characteristic function of the ball $bp^{-r}+p^{-r}\mathbb{Z}_{p}$.

We now define
\[
\psi_{rbk}\left(  x_{p}\right)  =p^{\frac{-r}{2}}\chi_{p}(p^{-1}k\left(
p^{r}x_{p}-b\right)  )\Omega\left(  \left\vert p^{r}x_{p}-b\right\vert
_{p}\right)  ,
\]
where $r\in\mathbb{Z}$, $k\in\{1,\dots,p-1\}$, and $b\in\mathbb{Q}%
_{p}/\mathbb{Z}_{p}$.

Then, $\left\{  \psi_{rbk}\left(  x_{p}\right)  \right\}  _{rbk}$ forms a
complete orthonormal basis of $L^{2}(\mathbb{Q}_{p})$, and
\[
\boldsymbol{D}^{\alpha}\psi_{rbk}\left(  x\right)  =p^{\left(  1-r\right)
\alpha}\psi_{rbk}\left(  x\right)  \text{, for any }r,b,k;
\]
see, e.g., \cite[Theorems 9.4.5 and 8.9.3]{Alberio et al}, or \cite[Theorem
3.3]{KKZuniga}

\subsection{Some results about tensor product of Hilbert spaces}

We set $L^{2}(\mathbb{R},dx_{\infty})=L^{2}(\mathbb{R})$, where $dx_{\infty}%
$\ denotes the Lebesgue measure, and $L^{2}\left(  \mathbb{Q}_{p}%
,dx_{p}\right)  =L^{2}\left(  \mathbb{Q}_{p}\right)  $, where $dx_{p}$ denotes
the Haar measure of $\mathbb{Q}_{p}$. We denote by $dx_{\infty}dx_{p}$, the
product measure of $dx_{\infty}$ and $dx_{p}$ on $\mathbb{R}\times
\mathbb{Q}_{p}$. On the other hand, $L^{2}\left(  \mathbb{Q}_{p}\right)  $\ is
a separable Hilbert space because it has a countable orthonormal basis
$\left\{  \psi_{rbk}\left(  x_{p}\right)  \right\}  _{rbk}$, and also
$L^{2}(\mathbb{R})$ is a separable Hilbert space. The Haar wavelets $\left\{
H_{m}(x_{\infty})\right\}  _{m\in\mathbb{N}}$, which are functions with
compact support, form an orthonormal basis of $L^{2}(\mathbb{R})$; also the
Hermite polynomials, which are Schwartz functions in $\mathbb{R}$, form an
orthonormal basis of $L^{2}(\mathbb{R})$. Then $\left\{  H_{m}\left(
x_{\infty}\right)  \psi_{rbk}\left(  x_{p}\right)  \right\}  _{rbk,m}$ is an
orthonormal basis of $L^{2}(\mathbb{R})\otimes L^{2}\left(  \mathbb{Q}%
_{p}\right)  $, cf. \cite[Proposition 2.7]{Asao} or \cite[Section II.4,
Proposition 2]{Reed-Simon-I}, where $\otimes$\ denotes the tensor product of
Hilbert spaces.

There is a unique isomorphism from $L^{2}(\mathbb{R},dx_{\infty})\otimes
L^{2}\left(  \mathbb{Q}_{p},dx_{p}\right)  $\ to $L^{2}\left(  \mathbb{R}%
\times\mathbb{Q}_{p},dx_{\infty}dx_{p}\right)  $ such that $f%
{\textstyle\bigotimes}
g$ is taken into $fg$; and there is a unique isomorphism from $L^{2}\left(
\mathbb{R}\times\mathbb{Q}_{p},dx_{\infty}dx_{p}\right)  $ to $L^{2}\left(
\mathbb{R},dx_{\infty};L^{2}\left(  \mathbb{Q}_{p},dx_{p}\right)  \right)  $
such that $f(x_{\infty},x_{p})$ is taken into the function $x_{p}\rightarrow
f(\cdot,x_{p})$. Similarly, there is a unique isomorphism from $L^{2}\left(
\mathbb{R}\times\mathbb{Q}_{p},dx_{\infty}dx_{p}\right)  $ to $L^{2}\left(
\mathbb{Q}_{p},dx_{p};L^{2}\left(  \mathbb{R},dx_{\infty}\right)  \right)  $
such that $f(x_{\infty},x_{p})$ is taken into the function $x_{\infty
}\rightarrow f(x_{\infty},\cdot)$; see \cite[Section II.4, Theorem
II.10]{Reed-Simon-I}.

\subsection{QM on $L^{2}(\mathbb{R\times Q}_{p})$}

Take $\Psi\left(  t\right)  =\Psi\left(  x_{\infty},x_{p},t\right)  \in
L^{2}(\mathbb{R})\otimes L^{2}\left(  \mathbb{Q}_{p}\right)  $. For a fixed
value of $x_{\infty}$, we set $\Psi\left(  x_{\infty},x_{p},t\right)
=\Psi_{p}\left(  x_{p},t\right)  $; and for a fixed value of $x_{p}$, we set
$\Psi\left(  x_{\infty},x_{p},t\right)  =\Psi_{\infty}\left(  x_{\infty
},t\right)  $. These functions satisfy $\Psi_{\infty}\left(  \cdot,t\right)
\in L^{2}(\mathbb{R})$, $\Psi_{p}\left(  \cdot,t\right)  \in L^{2}\left(
\mathbb{Q}_{p}\right)  $ for any $t\in\mathbb{R}$ .

Now, given a wavefunction $\Psi\left(  x_{\infty},x_{p},t\right)  $, the Born
rule produces several different probability distributions.

\textbf{Type 1}

We fix $x_{p}$, then%
\[
\frac{\left\vert \Psi\left(  x_{\infty},x_{p},t\right)  \right\vert ^{2}%
}{\left\Vert \Psi\left(  \cdot,x_{p},t\right)  \right\Vert _{L^{2}%
(\mathbb{R})}^{2}}dx_{\infty}=\frac{\Psi_{\infty}\left(  x_{\infty},t\right)
}{\left\Vert \Psi_{\infty}\left(  \cdot,t\right)  \right\Vert _{L^{2}%
(\mathbb{R})}^{2}}dx_{\infty}%
\]
is a probability measure\ in $\left(  \mathbb{R},\mathcal{B}(\mathbb{R}%
)\right)  $, for $t\in\mathbb{R}$, where $\mathcal{B}(\mathbb{R})$ is the
Borel $\sigma$-algebra of $\mathbb{R}$.

\textbf{Type 2}

We fix $x_{\infty}$, then%

\[
\frac{\left\vert \Psi\left(  x_{\infty},x_{p},t\right)  \right\vert ^{2}%
}{\left\Vert \Psi\left(  x_{\infty},\cdot,t\right)  \right\Vert _{L^{2}%
(\mathbb{Q}_{p})}^{2}}dx_{p}=\frac{\Psi_{p}\left(  x_{p},t\right)
}{\left\Vert \Psi_{p}\left(  \cdot,t\right)  \right\Vert _{L^{2}%
(\mathbb{Q}_{p})}^{2}}dx_{p}%
\]
is a probability measure\ in $\left(  \mathbb{Q}_{p},\mathcal{B}%
(\mathbb{Q}_{p})\right)  $, for $t\in\mathbb{R}$, where $\mathcal{B}%
(\mathbb{Q}_{p})$ is the Borel $\sigma$-algebra of $\mathbb{Q}_{p}$.

\textbf{Type 3}

For $t\in\mathbb{R}$,%
\[
\frac{\left\vert \Psi\left(  x_{\infty},x_{p},t\right)  \right\vert ^{2}%
}{\left\Vert \Psi\left(  \cdot,\cdot,t\right)  \right\Vert _{L^{2}%
(\mathbb{R\times Q}_{p})}^{2}}dx_{\infty}dx_{p}%
\]
is a probability measure\ in $\left(  \mathbb{R\times Q}_{p},\mathcal{B}%
(\mathbb{R\times Q}_{p})\right)  $, where $\mathcal{B}(\mathbb{R\times Q}%
_{p})$ is the Borel $\sigma$-algebra of $\mathbb{R\times Q}_{p}$.

There are three different quantum states, $\Psi\left(  x_{\infty}%
,x_{p},t\right)  $, $\Psi_{\infty}\left(  x_{\infty},t\right)  $, $\Psi
_{p}\left(  x_{p},t\right)  $; we argue that they provide, in principle, three
different pictures of the quantum reality. Notice, that in general,
$\Psi\left(  x_{\infty},x_{p},t\right)  \neq\Psi_{\infty}\left(  x_{\infty
},t\right)  \Psi_{p}\left(  x_{p},t\right)  $. We will come back to this
discussion in Section \ref{Section_7}.

\subsection{Non-local Schr\"{o}dinger equations}

The next step is to determine the unitary evolution operators acting on
$L^{2}(\mathbb{R})\otimes L^{2}\left(  \mathbb{Q}_{p}\right)  $, which may
have a physical meaning. We assume that the standard QM on $L^{2}\left(
\mathbb{R}^{N}\right)  $ provides a good description of quantum phenomena,
when $\mathbb{R}\times\mathbb{R}^{N}$ is a good model for the physical
configuration space associated with the phenomena. We also assume that
$p$-adic QM on $L^{2}\left(  \mathbb{Q}_{p}^{N}\right)  $ accurately describes
quantum phenomena, when $\mathbb{R}\times\mathbb{Q}_{p}^{N}$ is a good model
for the physical configuration space. By using a model of the form
$\mathbb{R}\times\left(  \mathbb{R}\times\mathbb{Q}_{p}\right)  ^{N}$, we have
regions of type%
\[
\mathbb{R}\times\left(  \mathbb{R}\times\left\{  \alpha\right\}  \right)
^{N}\simeq\mathbb{R}\times\mathbb{R}^{N},
\]
for some fixed $\alpha\in\mathbb{Q}_{p}$, where the standard (local) QM is
valid, and regions of the form%
\[
\mathbb{R}\times\left(  \left\{  \beta\right\}  \times\mathbb{Q}_{p}\right)
^{N}\simeq\mathbb{R}\times\mathbb{Q}_{p}^{N},
\]
for some fixed $\beta\in\mathbb{R}$, where $p$-adic (non-local) QM is valid.

Now, as before, for the sake of simplicity, we take $N=1$. The goal is to
describe the evolution operators $e^{it\boldsymbol{H}}:L^{2}\left(
\mathbb{R\times Q}_{p}\right)  \rightarrow L^{2}\left(  \mathbb{R\times Q}%
_{p}\right)  $ that satisfy the hypotheses formulated in the previous
paragraph. To determine the operator $\boldsymbol{H}$, it is sufficient to
describe it on a dense subset of $L^{2}\left(  \mathbb{R\times Q}_{p}\right)
$.\ The functions $\psi_{p}\left(  x_{p}\right)  \psi_{\infty}\left(
x_{\infty}\right)  $, with $\psi_{p}\left(  x_{p}\right)  \in L^{2}\left(
\mathbb{Q}_{p}\right)  $, $\psi_{\infty}\left(  x_{\infty}\right)  \in
L^{2}\left(  \mathbb{R}\right)  $, are dense in $L^{2}\left(  \mathbb{R\times
Q}_{p}\right)  $. For this reason, we take $\Psi\left(  x_{\infty}%
,x_{p},t\right)  =\Psi_{p}\left(  x_{p},t\right)  \Psi_{\infty}\left(
x_{\infty},t\right)  $, where $\Psi_{\infty}\left(  x_{\infty},t\right)  $ is
a quantum state evolving according to the standard QM on $\mathbb{R}%
\times\mathbb{R}$, and $\Psi_{p}\left(  x_{p},t\right)  $ is a quantum state
evolving according to the $p$-adic QM on $\mathbb{R}\times\mathbb{Q}_{p}$,
i.e.,%
\[
\left\{
\begin{array}
[c]{l}%
i\frac{\partial}{\partial t}\Psi_{\infty}\left(  x_{\infty},t\right)
=\boldsymbol{H}_{\infty}\Psi_{\infty}\left(  x_{\infty},t\right)  \text{,
\ }x_{\infty}\in\mathbb{R},t\geq0;\\
\\
\Psi_{\infty}\left(  \cdot,t\right)  \in L^{2}\left(  \mathbb{R}\right)
,\text{ }t\geq0;\\
\\
\left\Vert \Psi_{p}\left(  \cdot,t\right)  \right\Vert _{2}=1,\text{ }t\geq0,
\end{array}
\right.
\]
and
\[
\left\{
\begin{array}
[c]{l}%
i\frac{\partial}{\partial t}\Psi_{p}\left(  x_{p},t\right)  =\boldsymbol{H}%
_{p}\Psi_{p}\left(  x_{p},t\right)  \text{, \ }x_{p}\in\mathbb{Q}_{p}%
,t\geq0;\\
\\
\Psi_{p}\left(  \cdot,t\right)  \in L^{2}\left(  \mathbb{Q}_{p}\right)
,\text{ }t\geq0;\\
\\
\left\Vert \Psi_{p}\left(  \cdot,t\right)  \right\Vert _{2}=1,\text{ }t\geq0,
\end{array}
\right.
\]
where $\boldsymbol{H}_{p}:L^{2}\left(  \mathbb{Q}_{p}\right)  \supset
Dom(\boldsymbol{H}_{\infty})\rightarrow L^{2}\left(  \mathbb{Q}_{p}\right)  $,
and $\boldsymbol{H}_{\infty}:L^{2}\left(  \mathbb{R}\right)  \supset
Dom(\boldsymbol{H}_{\infty})\rightarrow L^{2}\left(  \mathbb{R}\right)  $ are
Hamiltonians, and $\boldsymbol{H}_{p}$ is a non-local operator, while
$\boldsymbol{H}_{\infty}$ is a local operator.

A straightforward calculation shows that $\Psi\left(  x_{\infty}%
,x_{p},t\right)  $ obeys to the following Schr\"{o}dinger equation:%
\begin{equation}
\left\{
\begin{array}
[c]{l}%
i\frac{\partial}{\partial t}\Psi\left(  x_{\infty},x_{p},t\right)  =\left(
\boldsymbol{H}_{\infty}+\boldsymbol{H}_{p}\right)  \Psi\left(  x_{\infty
},x_{p},t\right)  \text{, \ }\left(  x_{\infty},x_{p}\right)  \in
\mathbb{R\times Q}_{p},t\geq0;\\
\\
\Psi\left(  \cdot,\cdot,t\right)  \in L^{2}(\mathbb{R}\otimes\mathbb{Q}%
_{p}),\text{ }t\geq0;\\
\\
\left\Vert \Psi_{p}\left(  \cdot,t\right)  \right\Vert _{L^{2}(\mathbb{R}%
\otimes\mathbb{Q}_{p})}=1,\text{ }t\geq0,
\end{array}
\right.  \label{Non-local-Schrodinger}%
\end{equation}
where the Hamiltonian $\boldsymbol{H}_{\infty}+\boldsymbol{H}_{p}:L^{2}\left(
\mathbb{R}\right)
{\textstyle\bigotimes}
L^{2}\left(  \mathbb{Q}_{p}\right)  \supset Dom(\boldsymbol{H}_{\infty
}+\boldsymbol{H}_{p})\rightarrow L^{2}\left(  \mathbb{R}\right)
{\textstyle\bigotimes}
L^{2}\left(  \mathbb{Q}_{p}\right)  $ is a non-local operator.

Non-local operators of type $\boldsymbol{H}_{\infty}+\boldsymbol{H}_{p}$ (of
course with $\boldsymbol{H}_{p}\neq0$) allow spooky action at a
distance.\ Then, QM (in the sense of Dirac-von Neumann) on the Hilbert space
$L^{2}\left(  \mathbb{R}\right)
{\textstyle\bigotimes}
L^{2}\left(  \mathbb{Q}_{p}\right)  $, with time evolution $\left\{
e^{-it\left(  \boldsymbol{H}_{\infty}+\boldsymbol{H}_{p}\right)  }\right\}
_{t\in\mathbb{R}}$ is a non-local theory; consequently, the fact the universe
is not locally real, implies that this QM admits realism.

\begin{remark}
The above argument only gives a subset of possible admissible Hamiltonians for
our quantum formalism. Notice that any operator of the form%
\[
\boldsymbol{H}=\boldsymbol{H}_{\infty}+\boldsymbol{P}\left(  \boldsymbol{H}%
_{p},\boldsymbol{H}_{\infty}\right)  +\boldsymbol{H}_{p},
\]
where operator $\boldsymbol{P}$ satisfies%
\[
\boldsymbol{P}\left(  \boldsymbol{0},\boldsymbol{H}_{\infty}\right)
=\boldsymbol{P}\left(  \boldsymbol{H}_{p},\boldsymbol{0}\right)
=\boldsymbol{0}%
\]
should be considered as admissible.
\end{remark}

We now assume that $\left\{  \psi_{rbk}\left(  x_{p}\right)  \right\}  _{rbk}$
is an orthonormal basis of $L^{2}\left(  \mathbb{Q}_{p}\right)  $ consisting
of eigenfunctions of $\boldsymbol{H}_{p}$,%
\[
\boldsymbol{H}_{p}\psi_{rbk}=E_{rbk}\psi_{rbk}\text{, with }E_{rbk}%
\in\mathbb{R}\text{,}%
\]
and that $\left\{  \theta_{m}\left(  x_{\infty}\right)  \right\}  _{m}$ is an
orthonormal basis of $L^{2}\left(  \mathbb{R}\right)  $ consisting of
eigenfunctions of $\boldsymbol{H}_{\infty}$,%
\[
\boldsymbol{H}_{\infty}\theta_{m}=E_{m}\theta_{m}\text{, with }E_{m}%
\in\mathbb{R}\text{.}%
\]
Then, any solution of the Schr\"{o}dinger equation in
\ (\ref{Non-local-Schrodinger}) has the form%
\[
\Psi\left(  x_{\infty},x_{p},t\right)  =%
{\displaystyle\sum\limits_{rbk}}
{\displaystyle\sum\limits_{m}}
A_{rbk,m}e^{-it\left(  E_{rbk}+E_{m}\right)  }\theta_{m}\left(  x_{\infty
}\right)  \psi_{rbk}\left(  x_{p}\right)  .
\]
Since $\left\{  \theta_{m}\left(  x_{\infty}\right)  \psi_{rbk}\left(
x_{p}\right)  \right\}  _{rbk,m}$ is an orthonormal basis of $L^{2}%
(\mathbb{R})\otimes L^{2}\left(  \mathbb{Q}_{p}\right)  \simeq L^{2}%
(\mathbb{R}\otimes\mathbb{Q}_{p})$, $\Psi\left(  x_{\infty},x_{p},t\right)
\in L^{2}(\mathbb{R}\otimes\mathbb{Q}_{p})$, for any $t\geq0$. We now fix the
coefficients $A_{rbk,m}$ so that%
\[
\Psi\left(  x_{\infty},x_{p},0\right)  =%
{\displaystyle\sum\limits_{rbk}}
{\displaystyle\sum\limits_{m}}
A_{rbk,m}\theta_{m}\left(  x_{\infty}\right)  \psi_{rbk}\left(  x_{p}\right)
.
\]
Finally, using that $\left\{  \theta_{m}\left(  x_{\infty}\right)  \psi
_{rbk}\left(  x_{p}\right)  \right\}  _{rbk,m}$ is an orthonormal basis of
$L^{2}(\mathbb{R}\otimes\mathbb{Q}_{p})$,%
\[
\left\Vert \Psi\left(  x_{\infty},x_{p},t\right)  \right\Vert _{L^{2}%
(\mathbb{R}\otimes\mathbb{Q}_{p})}^{2}=%
{\displaystyle\sum\limits_{rbk}}
{\displaystyle\sum\limits_{m}}
\left\vert A_{rbk,m}\right\vert ^{2}=\left\Vert \Psi\left(  x_{\infty}%
,x_{p},0\right)  \right\Vert _{L^{2}(\mathbb{R}\otimes\mathbb{Q}_{p})}%
^{2}=1\text{,}%
\]
for $t\geq0$.

\section{\label{Section_5}The measurement problem and GWR\ theory}

In this section, we review quickly the measurement problem and the
Ghirardi-Rimini-Weber (GRW) collapse theory. In the next section, we propose a
solution to the measurement problem. The new collapse mechanism resembles the
one in the GRW theory, but there are several fundamental differences.

\subsection{The measurement problem}

We review the measurement problem following Bell, \cite{Bell -Againts-Meas},
and Landau and Lifshitz, \cite{Landau-Lifschitz}.

We consider a system consisting of two parts: a classical apparatus (a
macroscopic system) and a quantum object (a microscopic system). The process
of measurement involves the interaction of these two parts; as a result the
apparatus passes from its initial state into some other. From this change of
state one draws conclusions concerning the state of the quantum object. Let
$g$ be the pointer of the apparatus $\mathcal{A}$, i.e., the readings of it.
Let $\widehat{\boldsymbol{g}}$ be the self-adjoint operator corresponding to
the observable $g$, which we suppose to have a discrete and non-degenerate
spectrum. Let $g_{n}\in\mathbb{R}$, $F_{n}\in\mathcal{H}_{\mathcal{A}}$,
$n=1,2,\ldots$, be, respectively, the eigenvalues and eigenfunctions of
$\widehat{\boldsymbol{g}}$, where $\mathcal{H}_{\mathcal{A}}$\ is the state
space of $\mathcal{A}$. We emphasize that the wavefunctions of the apparatus
$F_{n}=F_{n}\left(  \xi,t\right)  $, $\left(  \xi,t\right)  \in\mathbb{R}%
^{N}\times\mathbb{R}$ are functions in the configuration space $\mathbb{R}%
^{N}\times\mathbb{R}$, and thus by our previous discussion the apparatus is in
the macroscopic realm. Let $F_{0}\left(  \xi\right)  =F_{n_{0}}\left(
\xi,0\right)  $ be the wavefunction of the initial state of the apparatus at
$t=0$ before the measurement.

Let $\psi_{n}$, $n=1,2,\ldots$ , be the eigenvectors of the operator
$\widehat{\boldsymbol{o}}$ corresponding to the observable $o$ that is to be
determined by the apparatus. We assume that $\Psi\left(  q,t\right)  =\sum
_{n}c_{n}\left(  t\right)  \psi_{n}$, $\left(  q,t\right)  \in\mathcal{X}%
^{N}\times\mathbb{R}$, \ meaning that each $\Psi\left(  q,t\right)  $ is the
state of a quantum object in $\mathcal{X}^{N}\times\mathbb{R}$. \ Only in this
section, $\mathcal{X}$ is an arbitrary space, so it can be $\mathbb{R}$ or
$\mathbb{Q}_{p}$.

Let%
\[
\Psi_{0}\left(  q\right)  =%
{\displaystyle\sum\limits_{n}}
c_{n}\left(  0\right)  \psi_{n}\left(  q\right)
\]
be the initial state of the quantum object. The initial state of the whole
system before the measurement is
\[
\Psi_{0}%
{\textstyle\bigotimes}
F_{0}=\Psi_{0}\left(  q\right)  F_{0}\left(  \xi\right)  \in\mathcal{H}%
_{\mathcal{S}}%
{\textstyle\bigotimes}
\mathcal{H}_{\mathcal{A}},
\]
where $\mathcal{H}_{\mathcal{S}}$\ is the Hilbert space corresponding to the
quantum object $\mathcal{S}$. Due to the interaction between $\mathcal{S}$ and
$\mathcal{A}$, this wavefunction evolves according the Schr\"{o}dinger
equation. We denote by $\Psi\left(  q,\xi,t\right)  $ the wavefunction of the
system $\mathcal{S}+\mathcal{A}$ at time $t$, then%
\begin{equation}
\Psi\left(  q,\xi,t\right)  =%
{\displaystyle\sum\limits_{n}}
a_{n}\psi_{n}\left(  q,t\right)  F_{n}\left(  \xi,t\right)  =%
{\displaystyle\sum\limits_{n}}
A_{n}\left(  q,t\right)  F_{n}\left(  \xi,t\right)  .\label{wavefunction}%
\end{equation}
The essence of the measurement problem\ is that after the interaction of
$\mathcal{S}$, and $\mathcal{A}$, the Schr\"{o}dinger equation does not give a
definite post-interaction position for the measuring apparatus pointer. In QM,
this problem is fixed by introducing the \textquotedblleft collapse
postulate\textquotedblright: when the measurement occurs, the wavefunction
ceases to evolve according to the Schr\"{o}dinger equation momentarily\ does
something entirely different.

Following Landau and Lifshitz, \cite{Landau-Lifschitz}, the classical nature
of the apparatus implies that at any instant $t>0$, the quantity $g$ (the
reading of the apparatus) has a definite value, and thus after the measurement
the system $\mathcal{S}+\mathcal{A}$ will be not described by
(\ref{wavefunction}), but by the term corresponding to the reading $g_{n}$ of
the apparatus,%
\[
A_{n}\left(  q,t\right)  F_{n}\left(  \xi,t\right)  .
\]
By writing
\[
A_{n}\left(  q,t\right)  =\left\Vert A_{n}\left(  \cdot,t\right)  \right\Vert
_{2}\frac{A_{n}\left(  q,t\right)  }{\left\Vert A_{n}\left(  \cdot,t\right)
\right\Vert _{2}},
\]
one gets that $\frac{A_{n}\left(  q,t\right)  }{\left\Vert A_{n}\left(
\cdot,t\right)  \right\Vert _{2}}$ is the normalized wavefunction of
$\mathcal{S}$ after the measurement.

\subsection{The GRW collapse model}

Ghirardi-Rimini-Weber (GRW) theory, and other spontaneous collapse theories,
solve the quantum measurement problem by modifying the Schr\"{o}\-dinger
equation with nonlinear and stochastic terms that cause the wavefunction to
collapse spontaneously. We give a quick review of the GRW collapse model; see,
e.g., \cite{Norsen}-\cite{Bell-q-Jumps}, and the references therein. This
section is based on \cite{Norsen} and \cite{Bell-q-Jumps}.

In the GRW proposal, the wavefunctions evolve according to the Schr\"{o}dinger
equation most of the time, except for occasional random moments when they
suffer a spontaneous localization. Consider the Gaussian function%
\[
g_{r}\left(  x\right)  =\frac{1}{\left(  2\pi\sigma^{2}\right)  ^{\frac{1}{4}%
}}e^{-\frac{\left(  x-r\right)  ^{2}}{4\sigma^{2}}}\text{, with }%
{\displaystyle\int\limits_{\mathbb{R}}}
g_{r}^{2}\left(  x\right)  dx=1,
\]
centered at the point $x=r$. During the spontaneous localization at time $t$,
the wavefunction gets multiplied by $g_{r}\left(  x\right)  $:%
\begin{equation}
\Psi\left(  x,t_{+}\right)  =\frac{g_{r}\left(  x\right)  \Psi\left(
x,t_{-}\right)  }{P(r)}, \label{Locaalization}%
\end{equation}
where $\Psi\left(  x,t_{-}\right)  $ is the wavefunction before time $t$,
$\Psi\left(  x,t_{+}\right)  $ is the wavefunction after time $t$, and
\[
P(r)=\sqrt{%
{\displaystyle\int\limits_{\mathbb{R}}}
\left\vert g_{r}\left(  x\right)  \Psi\left(  x,t_{-}\right)  \right\vert
^{2}dx}.
\]
The parameter $r$ is a random variable with a probability distribution $P(r)$.
The localization (\ref{Locaalization}) occurs at random times, with a Poisson
distribution with mean frequency $\lambda$. The parameters $\lambda$ and
$\sigma$ are phenomenological constants that should be understood as new
constants of Nature. The theory can be formulated through a short list of
principles, from these using the statistical operator formalism, a master
equation for $N$ particles can be obtained. This equation is a substitute to
the standard Schr\"{o}dinger equation; see \cite{Ghirardi-Romano}, and the
references therein.

It is relevant that $\Psi\left(  x,t_{-}\right)  \rightarrow\Psi\left(
x,t_{+}\right)  $ tends to localize $\Psi\left(  x,t_{-}\right)  $ because the
Gaussian function is concentrated around $x=r$ for a suitable choice of
$\sigma$. We invite the reader to review the several examples presented in
Norsen's book \cite[Chapter 9]{Norsen}. The collapse mechanism in the GWR
theory\ resembles the one introduced here. However, the perspective and the
implications of the GWR theory are radically different\ from those of our
theory, as we shall discuss in the coming sections.

\section{\label{Section_6}The collapse (localization) of the wavefunctions}

We now apply the quantum mechanical formalism developed in the previous
sections to propose a solution to the collapse of the wavefunction. In our
formalism, the collapse of the wavefunction is a consequence of the geometry
of the space-time (the configuration space).

We use the notation introduced at the end of Section \ref{Section_4}. The
apparatus is in an space-time that can be modeled as $\mathbb{R}%
\times\mathbb{R}$, where the locality and realism are valid; while the quantum
particle is in an space-time that can be modeled as $\mathbb{R}\times
\mathbb{Q}_{p}$, where non-locality and realism are valid. The wavefunctions
of the quantum particle are of the form $\Psi_{p}\left(  x_{p},t\right)  \in
L^{2}\left(  \mathbb{Q}_{p}\right)  $, for any $t\geq0$, and
\begin{equation}
\Psi_{p}\left(  x_{p},t\right)  =%
{\displaystyle\sum\limits_{rbk}}
c_{rbk}\left(  t\right)  \psi_{rbk}\left(  x_{p}\right)  ; \label{Exapnsion_P}%
\end{equation}
while the wavefunctions of the apparatus are of the form $\psi_{\infty}\left(
x_{\infty},t\right)  \in L^{2}\left(  \mathbb{R}\right)  $, for any $t\geq0$,
and%
\begin{equation}
\Psi_{\infty}\left(  x_{\infty},t\right)  =%
{\displaystyle\sum\limits_{m}}
c_{m}\left(  t\right)  \theta_{m}\left(  x_{\infty}\right)  .
\label{Exapnsion_Innity}%
\end{equation}
The wavefunction of the system $\mathcal{S}+\mathcal{A}$ at time $t$, is given
by%
\begin{equation}
\Psi\left(  x_{\infty},x_{p},t\right)  =%
{\displaystyle\sum\limits_{rbk}}
{\displaystyle\sum\limits_{m}}
a_{rbk,m}\left(  t\right)  \psi_{rbk}\left(  x_{p}\right)  \theta_{m}\left(
x_{\infty}\right)  , \label{wavefunction_2A}%
\end{equation}
where $a_{rbk,m}\in\mathbb{C}$. We do not assume that $\Psi\left(  x_{\infty
},x_{p},t\right)  $ is the product of $\Psi_{p}\left(  x_{p},t\right)  $\ and
$\Psi_{\infty}\left(  x_{\infty},t\right)  $. Then, (\ref{wavefunction_2A}) is
exactly (\ref{wavefunction}), when we use the space-time model $\mathbb{R}%
\times\left(  \mathbb{R}\times\mathbb{Q}_{p}\right)  $. The time evolution of
the quantum state $\Psi\left(  x_{\infty},x_{p},t\right)  $ obeys to a
non-local Schr\"{o}dinger equation on $\mathbb{R}\times\left(  \mathbb{R}%
\times\mathbb{Q}_{p}\right)  $.

\subsection{Key calculations}

The Monna map is defined as%
\[%
\begin{array}
[c]{cccc}%
\mathcal{M}: & \mathbb{Q}_{p} & \rightarrow & \mathbb{R}_{\geq0}\\
&  &  & \\
& x_{p}=%
{\displaystyle\sum\limits_{j=\gamma}^{\infty}}
y_{j}p^{j} & \rightarrow & x_{\infty}=%
{\displaystyle\sum\limits_{j=\gamma}^{\infty}}
y_{j}p^{-j-1}.
\end{array}
\]
This map captures the strangeness of QM: $\mathbb{R}$ does not contain a copy
of $\mathbb{Q}_{p}$ that preserves both the topology and algebraic structure
of $\mathbb{Q}_{p}$. More precisely, $\mathbb{R}$ is not isomorphic to
$\mathbb{Q}_{p}$ as local fields, \cite[Chpater I, Section 3]{Weil}. The Monna
map is a continuous, surjective, but not injective, \cite[Section
1.9.4]{Alberio et al}.

At the measurement time, the apparatus scan a compact region in the
microscopic realm ($\mathbb{Q}_{p}$), the description of the classical
interaction between the apparatus and the quantum particle requires the
identification of points of $\mathbb{Q}_{p}$ with points in $\mathbb{R}$. A
such identification is obtained through the Monna map.

\subsubsection{\textbf{Formula 1}}

For each $t\in\mathbb{R}$, $\frac{1}{A\left(  \mathcal{M}\right)  }\left\vert
\Psi\left(  x_{\infty},x_{p},t\right)  \right\vert ^{2}dx_{p}$, with $x_{p}\in
B$, and $x_{\infty}=\mathcal{M}(x_{p})$ is the probability density, so%
\begin{equation}
P_{\text{int}}(B,t)=\frac{1}{A\left(  \mathcal{M}\right)  }%
{\displaystyle\int\limits_{B}}
\text{\ }\left\vert \Psi\left(  \mathcal{M}(x_{p}),x_{p},t\right)  \right\vert
^{2}dx_{p}\nonumber
\end{equation}
gives the probability of interaction between the apparatus and the quantum
object, when the apparatus scans a region $B\subset\mathbb{Q}_{p}$ at the the
time $t$. 

\subsubsection{\textbf{Formula 2}}

We assume that the scanned region $B$ is a ball, $B=p^{l}a+p^{l}\mathbb{Z}%
_{p}$, for $l\in\mathbb{N}$, $a\in\mathbb{Q}_{p}/\mathbb{Z}_{p}$. Then%
\[
\mathcal{M}\left(  B\right)  =\mathcal{M}(p^{l}a)+\left[  0,p^{-l}\right]
\]
is an interval in $\mathbb{R}_{\geq0}$; see \cite[Lemma 1.9.11]{Alberio et al}.

We now use expansion (\ref{Exapnsion_P}), where the support of $\left\{
\psi_{rbk}\right\}  _{rbk}$ (denoted as \textrm{supp} $\psi_{rbk}$) is a ball.
We now recall that in $p$-adic topology two balls are disjoint or one in
contained in the other; consequently, there are two cases:%
\begin{equation}
\mathrm{supp}\psi_{rbk}\subset B\text{ or }B\subset\mathrm{supp}\text{ }%
\psi_{rbk} \label{CASEA}%
\end{equation}
or%
\begin{equation}
\mathrm{supp}\text{ }\psi_{rbk}\cap B=\varnothing\Leftrightarrow
\mathrm{supp}\text{ }\psi_{rbk}\subset B^{c}, \label{CASEB}%
\end{equation}
where $B^{c}=\mathbb{Q}_{p}\smallsetminus B$. In case (\ref{CASEA}),
$\mathrm{supp}$ $\psi_{rbk}\cap B\neq\varnothing$. Using this fact, we rewrite
(\ref{Exapnsion_P}) as
\begin{equation}
\Psi_{p}\left(  x_{p},t\right)  =\Psi_{p}^{\left(  B\right)  }\left(
x_{p},t\right)  +\Psi_{p}^{\left(  B^{c}\right)  }\left(  x_{p},t\right)  ,
\label{FORMULA_PSI_1}%
\end{equation}
where%
\begin{equation}
\Psi_{p}^{\left(  B\right)  }\left(  x_{p},t\right)  :=%
{\displaystyle\sum\limits_{\substack{rbk\\\mathrm{supp}\text{ }\psi_{rbk}\cap
B\neq\varnothing}}}
c_{rbk}(t)\psi_{rbk}\left(  x_{p}\right)  \in\mathcal{H}_{\mathcal{S}},
\label{FORMULA_PSI_2}%
\end{equation}
and
\begin{equation}
\Psi_{p}^{\left(  B^{c}\right)  }\left(  x_{p},t\right)  =%
{\displaystyle\sum\limits_{\substack{rbk\\\mathrm{supp}\text{ }\psi_{rbk}\cap
B=\varnothing}}}
c_{rbk}(t)\psi_{rbk}\left(  x_{p}\right)  \in\mathcal{H}_{\mathcal{S}}.
\label{FORMULA_PSI_3}%
\end{equation}
Furthermore, $\left\langle \Psi_{p}^{\left(  B\right)  }\left(  x_{p}%
,t\right)  ,\Psi_{p}^{\left(  B^{c}\right)  }\left(  x_{p},t\right)  \text{
}\right\rangle =0$, for any $t\geq0$, and
\[
1_{B}\left(  x_{p}\right)  \Psi_{p}\left(  x_{p},t\right)  =\Psi_{p}^{\left(
B\right)  }\left(  x_{p},t\right)  .
\]
Therefore,%
\[%
{\displaystyle\int\limits_{B}}
\left\vert \Psi_{p}\left(  x_{p},t\right)  \right\vert ^{2}dx_{p}=%
{\displaystyle\int\limits_{\mathbb{Q}_{p}}}
\left\vert \Psi_{p}^{\left(  B\right)  }\left(  x_{p},t\right)  \right\vert
^{2}dx_{p}.
\]
In conclusion, the multiplication by $1_{B}\left(  x_{p}\right)  $ induces the
map%
\begin{equation}
\Psi_{p}\left(  x_{p},t\right)  =\Psi_{p}^{\left(  B\right)  }\left(
x_{p},t\right)  +\Psi_{p}^{\left(  B^{c}\right)  }\left(  x_{p},t\right)
\rightarrow\Psi_{p}^{\left(  B\right)  }\left(  x_{p},t\right)  ,
\label{Key_formula_2}%
\end{equation}
where $\Psi_{p}^{\left(  B\right)  }\left(  x_{p},t\right)  $ satisfies
(\ref{Key_formula_2}). During the scanning process, the apparatus localized
the wave function, $\Psi_{p}\left(  x_{p},t\right)  $, in the region
$B\subset\mathbb{Q}_{p}$. It is essential that the localization process
extract exactly the part of the wavefunction that describes the probability
density for the particle to be in $B$ at any time.

\subsubsection{\textbf{Formula 3}}

We now compute the restriction of $\psi_{rbk}\left(  x_{p}\right)  $ to the
ball $B=p^{l}a+p^{l}\mathbb{Z}_{p}$:
\begin{equation}
\Omega\left(  p^{l}\left\vert x_{p}-p^{l}a\right\vert _{p}\right)  \psi
_{rbk}\left(  x_{p}\right)  =\left\{
\begin{array}
[c]{ll}%
\psi_{rbk}\left(  x_{p}\right)  & \text{if }bp^{-r}-p^{l}a\in p^{l}%
\mathbb{Z}_{p}\text{, }r\leq-l\\
& \\
p^{\frac{-r}{2}}\Omega\left(  p^{l}\left\vert x_{p}-p^{l}a\right\vert
_{p}\right)  & \text{if }bp^{-r}-p^{l}a\in p^{-r}\mathbb{Z}_{p}\text{, }%
r\geq-l+1\\
& \\
0 & \text{if }bp^{-r}-p^{l}a\notin p^{-r}\mathbb{Z}_{p}\text{, }r\geq-l+1,
\end{array}
\right.  \label{TableA}%
\end{equation}
where $\Omega\left(  p^{l}\left\vert x_{p}-p^{l}a\right\vert _{p}\right)  $
denotes the characteristic function of $p^{l}a+p^{l}\mathbb{Z}_{p}$. The above
calculation has appeared in several publications; see, e.g., \cite[Table
4.4]{Zuniga-PhA}.

Now, using (\ref{wavefunction_2A}), and (\ref{TableA}),
\begin{gather}
\Omega\left(  p^{l}\left\vert x_{p}-p^{l}a\right\vert _{p}\right)  \Psi\left(
\mathcal{M}\left(  x_{p}\right)  ,x_{p},t\right)  =\label{Key_formula}\\%
{\displaystyle\sum\limits_{rbk}}
{\displaystyle\sum\limits_{m}}
a_{rbk,m}\left(  t\right)  \left\{  \Omega\left(  p^{l}\left\vert x_{p}%
-p^{l}a\right\vert _{p}\right)  \right\}  \psi_{rbk}\left(  x_{p}\right)
\theta_{m}\left(  \mathcal{M}\left(  x_{p}\right)  \right) \nonumber\\
=%
{\displaystyle\sum\limits_{r<-l}}
\ \
{\displaystyle\sum\limits_{b\in p^{r+l}b+p^{l+r}\mathbb{Z}_{p}}}
\
{\displaystyle\sum\limits_{m}}
a_{rbk,m}\left(  t\right)  \psi_{rbk}\left(  x_{p}\right)  \theta_{m}\left(
\mathcal{M}\left(  x_{p}\right)  \right)  +\nonumber\\
\Omega\left(  p^{l}\left\vert x_{p}-p^{l}a\right\vert _{p}\right)  \left\{
{\displaystyle\sum\limits_{r\geq-l+1}}
\
{\displaystyle\sum\limits_{b\in p^{r+l}b+p^{l+r}\mathbb{Z}_{p}}}
\
{\displaystyle\sum\limits_{m}}
p^{\frac{-r}{2}}a_{rbk,m}\left(  t\right)  \theta_{m}\left(  \mathcal{M}%
\left(  x_{p}\right)  \right)  \right\}  .\nonumber
\end{gather}

\subsection{The collapse mechanism}

By a localized wavefunction, we mean \ a compactly supported function in
$L^{2}$. The existence of compactly supported wavefunctions $\Psi_{\infty
}\left(  \cdot,t\right)  \in L^{2}\left(  \mathbb{R}^{N}\right)  $, $t\geq0$,
has a profound physical implications. For instance, the Dirac Hamiltonian does
not admit localized wavefunctions: any wavefunction with positive energy has
to be spread over all space ($\mathbb{R}^{3}$) at all times; see \cite[
Corollary 1.7]{Thaller}. The falsity of this result implies the violation of
Einstein's causality; see \cite[Section 1.8.2]{Thaller}.

We now discuss the localization (or collapse) of wavefunctions using the
formulas (\ref{Probability}) and (\ref{Key_formula}). At the time of the
measurement, the apparatus scans a small region of the microscopic space
($\mathbb{Q}_{p}$), which we assume to be a small ball $p^{l}a+p^{l}%
\mathbb{Z}_{p}$, where $l$ is a positive integer and $a\in\mathbb{Q}%
_{p}/\mathbb{Z}_{p}$. The ball $p^{l}a+p^{l}\mathbb{Z}_{p}$ corresponds to a
small interval $\mathcal{M}\left(  p^{l}a+p^{l}\mathbb{Z}_{p}\right)
=\mathcal{M}(p^{l}a)+\left[  0,p^{-l}\right]  $, in the macroscopic space
($\mathbb{R}$).

The scanning causes a classical interaction between the apparatus
$\mathcal{A}$ and the quantum object $\mathcal{S}$, which induces a
localization of the\ wavefunction of $\mathcal{S}$, i.e., it passes from
$\Psi_{p}\left(  x_{p},t\right)  $ to $\Psi_{p}\left(  x_{p},t\right)
\Omega\left(  p^{l}\left\vert x-p^{l}a\right\vert _{2}\right)  $, see
(\ref{Key_formula_2}). In turn, this localization induces a localization of
the wavefunction of the system $\mathcal{S}+\mathcal{A}$ around a small
interval, see (\ref{Key_formula}). We can symbolically describe the
localization of the wavefunction of the system as%
\begin{multline*}
\Psi\left(  x_{\infty},x_{p},t\right)  \rightarrow\text{ \ }\\
\Omega\left(  p^{l}\left\vert x-p^{l}a\right\vert _{2}\right)  \Psi\left(
\mathcal{M}\left(  x_{p}\right)  ,x_{p},t\right)  =\Omega\left(
p^{l}\left\vert x_{p}-p^{l}a\right\vert _{p}\right)
{\displaystyle\sum\limits_{m}}
A_{m}\left(  x_{p},t\right)  \theta_{m}\left(  \mathcal{M}\left(
x_{p}\right)  \right)  \text{, }%
\end{multline*}
where
\[
A_{m}\left(  x_{p},t\right)  =%
{\displaystyle\sum\limits_{r<-l}}
\ \
{\displaystyle\sum\limits_{b\in p^{r+l}b+p^{l+r}\mathbb{Z}_{p}}}
\ a_{rbk,m}\left(  t\right)  \psi_{rbk}\left(  x_{p}\right)  +%
{\displaystyle\sum\limits_{r\geq-l+1}}
\
{\displaystyle\sum\limits_{b\in p^{r+l}b+p^{l+r}\mathbb{Z}_{p}}}
\ p^{\frac{-r}{2}}a_{rbk,m}\left(  t\right)  ,
\]
for $x_{p}\in p^{l}a+p^{l}\mathbb{Z}_{p}$, see (\ref{Key_formula}).

There are infinitely many Hamiltonians $\boldsymbol{H}_{p}$, which are
(non-local) pseudo-differential operators, admitting eigenfunctions
$\psi_{rbk}\left(  x\right)  $\ supported in the ball $ap^{r}+p^{r}%
\mathbb{Z}_{p}$, for $a\in\mathbb{Q}_{p}/\mathbb{Z}_{p}$, $r\in\mathbb{Z}$.
Then, the proposed localization mechanism for the wavefunctions $\Psi\left(
x_{\infty},x_{p},t\right)  $ is mathematically realizable.

\subsection{An additional remark}

The collapse mechanism proposed does not work if we replace $\mathbb{R}%
\times\left(  \mathbb{R}\times\mathbb{Q}_{p}\right)  $ by $\mathbb{R}%
\times\left(  \mathbb{R}\times\mathbb{R}\right)  $.\ The calculations show in
Table \ref{TableA} use the fact that in $p$-adic topology, two balls are
disjoint or one is contained in the other. Using that $\mathrm{supp}\psi
_{rbk}=p^{-r}b+p^{-r}\mathbb{Z}_{p}$, the first line in Table \ref{TableA}
corresponds to the case $\mathrm{supp}\psi_{rbk}\subset B$; the second line
corresponds to $B\subset\mathrm{supp}$ $\psi_{rbk}$, and the last line
corresponds to $\mathrm{supp}$ $\psi_{rbk}\cap B=\varnothing$. This result is
not valid in the standard topology of $\mathbb{R}$. Based on this fact, we
assert that the proposed collapse mechanism is not valid in $\mathbb{R}%
\times\left(  \mathbb{R}\times\mathbb{R}\right)  $.

\subsection{A comparison with the GWR theory}

The GRW theory posits that the collapse of the wavefunction occurs in position
space, meaning that the wave function localizes to a specific region in space.
The wavefunctions are physical entities that are subject to a spontaneous and
random collapsing process in nature. The generalization of the GWR theory is
the QM with spontaneous localization; this theory is an alternative to the
standard QM.

In our proposal, the probability of interaction between the apparatus and the
quantum particle (computed via the Born rule in the microscopic space-time
$\mathbb{R}\times\mathbb{Q}_{p}$) has a density function compactly supported.
We argue that this fact can be interpreted as that the wavefunction of the
apparatus localizes during the measurement. From a mathematical perspective,
the described localization process is similar to the one given in the GWR
theory. However, in our framework, the wavefunctions are not physical
entities; only the probability measures that they define have physical
meaning. It is not necessary to introduce new physical constants, nor replace
the Schr\"{o}dinger equation with another equation involving nonlinear and
stochastic terms. In our framework, the collapse of the wavefunction is a
consequence of the difference between the geometry of the macroscopic realm
and the geometry of the microscopic one.

It is believed that GWR theory is compatible with the experimental evidence of
non-locality in the universe (the violation of Bell's inequalities). Since the
theory is formulated using the space-time model $\mathbb{R}\times
\mathbb{R}^{3}$, the compatibility with special relativity is a significant
problem. Finally, the quantum paradigm that the universe is not locally real
demands that we choose between realism and non-locality. The bottom line
question is whether GWR theory gives us a choice between (1) the universe is
not local, or (2) the universe is real. We argue that the GWR theory cannot
answer this question because it does not aim to explain the origin of
non-locality in QM, nor does it model non-local processes in QM; instead, GWR
theory uses non-locality only to explain the collapse of the wavefunction.

GRW theory establishes a natural collapse mechanism for wavefunctions, which
does not require an external observer, as pointed out in the Copenhagen
interpretation. Similarly, in our framework, the collapse of the wavefunction
is a consequence of the geometry of the configuration space.

\section{\label{Section_7}Two-slit Experiment in Quantum Mechanics}

In this section, we apply the quantum formalism developed to the two-slit
experiment. In \cite{Aharonov et al}, the authors provide a model of the
two-slit experiment without using the wave properties of the particle:
\textquotedblleft Instead of a quantum wave passing through both slits, we
have a localized particle with nonlocal interactions with the other
slit.\textquotedblright\ In \cite{Zuniga-AP}, the author proposes a $p$-adic
model for the same experiment, giving a similar conclusion, where the
non-local interactions are a consequence of the discreteness of the space
$\mathbb{Q}_{p}$.

In \cite{Zuniga-AP}, the evolution operators have the form
$e^{-it\boldsymbol{D}^{\alpha}}$, $\alpha>0$, for $t\in\mathbb{R}$, where
$\boldsymbol{D}^{\alpha}$ is non-local Taibleson-Vladimirov derivative; see
(\ref{Vladimirov-Taibleson-Derivative}). These operators are obtained from a
Feller semigroup $e^{-t\boldsymbol{D}^{\alpha}}$, $t\in\mathbb{R}_{+}$, by a
Wick rotation. The Feller feature means that there is a Markov process
attached to the semigroup with state space $\mathbb{Q}_{p}$. Intuitively, this
stochastic process describes a particle performing a random motion consisting
of jumps between the points of a self-similar set; see \cite[Chapter
2]{Zuniga-Textbook}, \cite[Chapter 2]{Zuniga-LNM-2016}. The free
Schr\"{o}dinger equation in natural units has the form $i\frac{\partial
\Psi_{p}\left(  x_{p},t\right)  }{\partial t}=\boldsymbol{D}^{\alpha}%
\Psi\left(  x_{p},t\right)  $, $x_{p}\in\mathbb{Q}_{p},t\geq0$. The function
$\left\vert \Psi\left(  x,t\right)  \right\vert ^{2}$ is a time-dependent
probability density, and $\int_{B}\left\vert \Psi\left(  x,t\right)
\right\vert ^{2}d^{N}x$ is the probability of finding a particle in the set
$B$ at the time $t$.

The mathematical setup for the ($p$-adic) double-slit experiment is as
follows. At time zero, there are two localized particles, which correspond to
two localized Gaussian \ probability densities at two different points $a$,
$b\in\mathbb{Q}_{p}$, such $\left\vert a-b\right\vert _{p}>p^{-L}$.
Geometrically, each slit corresponds to a ball centered at $a$ or $b$ with
radius $p^{-L}$, $L\geq0$, which is the size of each slit. These localized
particles evolve in time under a non-local evolution operator
$e^{-it\boldsymbol{D}^{\alpha}}$, $t\geq0$; the non-locality is a consequence
of the discreteness of space $\mathbb{Q}_{p}$. In standard QM, the fact that
the de Broglie waves (the quantum waves) are considered physical waves
explains the double-slit experiment as an interference phenomenon.

In the $p$-adic framework, de Broglie waves (`the quantum waves') are not
physical entities; thus, the explanation of the double-slit experiment as an
interference phenomenon is ruled out. However, the probability density
$\left\vert \Psi_{p}\left(  x_{p},t\right)  \right\vert ^{2}$ exhibits the
classical interference patterns attributed to `quantum waves.' In
\cite{Zuniga-AP}, we argue that the standard notion of trajectory, jointly
with the fact that the de Broglie waves are just mathematical objects,
concludes that in the double-slit experiment, each particle goes only through
one of the slits.

The model for the two-slit experiment given here is based on the models given
in\ \cite{Zuniga-AP} (case $\mathbb{Q}_{p}$), and \cite{Webb} (case
$\mathbb{R}$). We fix a non-zero rational number $s$, notice that
$s\in\mathbb{Q}_{p}$ and $s\in\mathbb{R}$, and set%
\[
\Psi_{p}^{\left(  0\right)  }\left(  x_{p}\right)  =A_{p}\left\{
\Omega\left(  p^{L}\left\vert x_{p}-s\right\vert _{p}\right)  +\Omega\left(
p^{L}\left\vert x_{p}+s\right\vert _{p}\right)  \right\}  ,
\]
where $L$ is a positive integer, and $A_{p}$ is a normalization constant such
that $\left\Vert \Psi_{p}^{\left(  0\right)  }\left(  x_{p}\right)
\right\Vert _{2}=1$, and%
\[
\Psi_{\infty}^{\left(  0\right)  }\left(  x_{\infty}\right)  =A_{\infty
}\left\{  e^{-\frac{\left(  x-s\right)  ^{2}}{2\sigma^{2}}}+e^{-\frac{\left(
x+s\right)  ^{2}}{2\sigma^{2}}}\right\}  ,
\]
where $\sigma$ is a positive constant, and $A_{\infty}$ is a normalization
constant such that $\left\Vert \Psi_{\infty}^{\left(  0\right)  }\left(
x_{\infty}\right)  \right\Vert _{2}=1$. Notice that $\Psi_{\infty}^{\left(
0\right)  }\left(  x_{\infty}\right)  $, $\Psi_{p}^{\left(  0\right)  }\left(
x_{p}\right)  $ are the superposition of two Gaussian packets located at $\pm
s$. We propose the following model for the two-slit experiment:
\begin{equation}
\left\{
\begin{array}
[c]{l}%
i\frac{\partial}{\partial t}\Psi\left(  x_{\infty},x_{p},t\right)  =\left\{
\frac{1}{2m_{p}}\boldsymbol{D}^{\alpha}-\frac{1}{2m_{\infty}}\frac
{\partial^{2}}{\partial x_{\infty}}\right\}  \Psi\left(  x_{\infty}%
,x_{p},t\right)  ;\\
\\
\Psi\left(  x_{\infty},x_{p},0\right)  =\Psi_{p}^{\left(  0\right)  }\left(
x_{p}\right)  \Psi_{\infty}^{\left(  0\right)  }\left(  x_{\infty}\right)  ;\\
\\
\left\Vert \Psi\left(  x_{\infty},x_{p},t\right)  \right\Vert _{L^{2}\left(
\mathbb{R}\times\mathbb{Q}_{p}\right)  }=1\text{, }%
\end{array}
\right.  \label{New-Model}%
\end{equation}
for $x_{p}\in\mathbb{Q}_{p}$, $x_{\infty}\in\mathbb{R}$, $t\geq0$. This
problem has a solution of the form $\Psi\left(  x_{\infty},x_{p},t\right)
=\Psi_{\infty}\left(  x_{\infty},t\right)  \Psi_{p}\left(  x_{p},t\right)  $,
where%
\begin{equation}
\left\{
\begin{array}
[c]{l}%
i\frac{\partial}{\partial t}\Psi_{\infty}\left(  x_{\infty},t\right)
=-\frac{1}{2m_{\infty}}\frac{\partial^{2}}{\partial x_{\infty}}\Psi_{\infty
}\left(  x_{\infty},t\right)  ;\\
\\
\Psi_{\infty}\left(  x_{\infty},0\right)  =\Psi_{\infty}^{\left(  0\right)
}\left(  x_{\infty}\right)  ;\\
\\
\left\Vert \Psi\left(  x_{\infty},t\right)  \right\Vert _{L^{2}\left(
\mathbb{R}\right)  }=1\text{, }%
\end{array}
\right.  \label{Archimedean-Moddel}%
\end{equation}
for $x_{\infty}\in\mathbb{R}$, $t\geq0$, and%
\begin{equation}
\left\{
\begin{array}
[c]{l}%
i\frac{\partial}{\partial t}\Psi_{p}\left(  x_{p},t\right)  =\frac{1}{2m_{p}%
}\boldsymbol{D}^{\alpha}\Psi_{p}\left(  x_{p},t\right)  ;\\
\\
\Psi_{p}\left(  x_{p},0\right)  =\Psi_{p}^{\left(  0\right)  }\left(
x_{p}\right)  ;\\
\\
\left\Vert \Psi_{p}\left(  x_{p},t\right)  \right\Vert _{L^{2}\left(
\mathbb{Q}_{p}\right)  }=1\text{, }%
\end{array}
\right.  \label{Non-Archimdean-Model}%
\end{equation}
where $x_{p}\in\mathbb{Q}_{p}$, $t\geq0$. The model (\ref{Archimedean-Moddel})
provides a good approximation of the interference patterns for the two-slit
experiment; see \cite{Webb}, and the references therein.\ The model
(\ref{Non-Archimdean-Model}) is a $p$-adic analogue of
(\ref{Archimedean-Moddel}), which gives interference patterns similar to the
standard ones, and provides a picture similar to the one given in
\cite{Aharonov et al}. Based on these reasons, we propose the model
(\ref{New-Model}).

There are two different quantum states $\Psi_{\infty}\left(  x_{\infty
},t\right)  $, $\Psi_{p}\left(  x_{p},t\right)  $ that produce two different
interference patterns, one in the macroscopic realm $\left\vert \Psi_{\infty
}\left(  x_{\infty},t\right)  \right\vert ^{2}$, and the other $\left\vert
\Psi_{p}\left(  x_{p},t\right)  \right\vert ^{2}$ in the microscopic one.
Here, it is important to remember that these two realms are connected via the
map $\mathcal{M}:$ $\mathbb{Q}_{p}\rightarrow\mathbb{R}$. We argue that the
pattern $\left\vert \Psi_{\infty}\left(  x_{\infty},t\right)  \right\vert
^{2}$ can be detected by an apparatus located in $\mathbb{R\times R}$, while
the pattern $\left\vert \Psi_{p}\left(  x_{p},t\right)  \right\vert ^{2}$
remains hidden for the apparatus.

In \cite{Villas-Boas et al}, the authors propose that the classical
interference patterns observed in experiments like the double-slit experiment
originate from collective \textquotedblleft bright\textquotedblright\ and
\textquotedblleft dark\textquotedblright\ states of light at the quantum
level. The bright states are quantum states of light that can readily interact
with a detector, while dark states are quantum states of light that cannot
interact with the detector. In this framework, it is proposed that photons
exist in these \textquotedblleft dark\textquotedblright\ regions of the
interference patterns. This proposal matches our previous discussion, if we
interpret $\Psi_{\infty}\left(  x_{\infty},t\right)  $ as a bright state, and
$\Psi_{p}\left(  x_{p},t\right)  $ as a dark state. However, from the
perspective of the quantum formalism introduced here, we expect that the
existence of dark states is a general fact in QM.

\section{\label{Section_8}Conclusions and final discussion}

QM on $\mathbb{C}^{N}$ is a particular case of $p$-adic QM, more precisely, it
is QM on $L^{2}(\chi_{N}\left(  \mathbb{Z}_{p}\right)  )$. $p$-Adic QM is
non-local because the Hamiltonians (non-local operators) allow spooky action
at a distance. Thus, this theory admits realism. The implementation of the
space discreteness hypothesis requires using a totally disconnected space
$\mathcal{X}$ as a model of the physical space at the microscopic level. In
such a space, there are no continuous curves (world lines), so any theory
using this type of space is not compatible with the theory of relativity. It
does not imply any immediate prediction on the violation of Einstein's causality.

We pick $\mathcal{X}=\mathbb{Q}_{p}$ because $p$-adic QM is the framework to
formulate the non-locality tests and to formulate continuous-time quantum
walks, which are a fundamental tool in quantum computing. This choice does not
discard that other totally disconnected spaces can be more convenient in
particular applications. Here, we use the $\mathbb{R}\times\left(
\mathbb{R}\times\mathbb{Q}_{p}\right)  ^{3}$ as a space-time model. Notice
that passing from $\mathbb{R}\times\mathbb{R}^{3}$ to $\mathbb{R}\times\left(
\mathbb{R}\times\mathbb{Q}_{p}\right)  ^{3}$ demands passing from
$4$-dimensions to $7$-dimensions. This larger space contains regions of the
form $\mathbb{R}\times\mathbb{R}^{3}$, where relativistic QM\ is valid; this
theory is local (uses local Hamiltonians) and real. At the macroscopic level,
the realism is not in dispute. Also, the mentioned space contains regions of
the form $\mathbb{R}\times\mathbb{Q}_{p}^{3}$,\ where $p$-adic QM is valid.

QM on the Hilbert space $L^{2}(\left(  \mathbb{R}\times\mathbb{Q}_{p}\right)
^{3})$, with evolution operators of the form $e^{it\left(  \boldsymbol{H}%
_{p}+\boldsymbol{H}_{\infty}\right)  }$ unifies the QM on $L^{2}%
(\mathbb{R}^{3})$ and $L^{2}(\mathbb{Q}_{p}^{3})$. In this framework, we
propose a collapse mechanism for the wavefunction that gives an alternative
solution to the measurement problem. In our proposal, the Schr\"{o}dinger
equation describes the evolution of the wavefunctions over time and their collapse.

The space-time model $\mathbb{R}\times\left(  \mathbb{R}\times\mathbb{Q}%
_{p}\right)  ^{3}$ allows breaking the causality and the Lorentz symmetry
naturally, when passing from the macroscopic space $\mathbb{R}^{3}$\ to the
microscopic one $\mathbb{Q}_{p}^{3}$. The Lorentz symmetry is one of the most
essential symmetries of the quantum field theory. While the validity of this
symmetry continues to be verified with a high degree of precision
\cite{Kostelecky-Russell}, in the last thirty-five years, the experimental and
theoretical studies of the Lorentz breaking symmetry have been an area of
intense research, see, e.g., \cite{Mariz et al}-\cite{Amelino-Camelia} and the
references therein.

In \cite{Zuniga-PhA}, we introduced a $p$-adic Dirac equation that shares many
properties with the standard one. In particular, the new equation also
predicts the existence of pairs of particles and antiparticles and a charge
conjugation symmetry. The geometry of space $\mathbb{Q}_{p}^{3}$ imposes
substantial restrictions on the solutions of the $p$-adic Dirac equation, \ in
particular, it admits space-localized planes waves varying in time. This
phenomenon does not occur in the standard case; see, e.g., \cite[Section 1.8,
Corollary 1.7]{Thaller}. On the other hand, we compute the transition
probability from a localized state at time $t=0$ to another localized state at
$t>0$, assuming that the space supports of the states are arbitrarily far
away. It turns out that this transition probability is greater than zero for
any time $t\in\left(  0,\epsilon\right)  $, for arbitrarily small $\epsilon$;
see \cite[Theorem 9.1]{Zuniga-PhA}. Since this probability is nonzero for some
arbitrarily small $t$, the system has a nonzero probability of getting between
the mentioned localized states arbitrarily shortly, thereby propagating with
superluminal speed in $\mathbb{R}\times\mathbb{Q}_{p}^{3}$.

Localization plays a central role in quantum field theory (QFT), and it is
itself an area of intense research; see \cite{Falcone-Conti}, and the
references therein. Here, we point out that the geometry of space-time
strongly influences localization. Using $\mathbb{R}\times\mathbb{R}^{3}$ as a
model of space-time, Newton and Wigner studied the localization of particles
in relativistic QM,\ \cite{Newton-Wigner}. The Newton-Wigner scheme in QFT
predicts a phenomenon of superluminal spreading, \cite{Felimng et al}, that
contradicts the relativistic notion of causality. This result is a consequence
of the Hegerfeldt theorem \cite{Hegerfeldt}, \cite{Hegerfeldt et al}, whose
only hypotheses are the positivity of the energy of relativistic particles and
the orthogonality condition of states localized in disjoint regions. The
Newton-Wigner scheme contradicts the no-communication theorem; for this
reason, it is not regarded as fundamental.

In 1988, Eberhard and Ross, using $\mathbb{R}\times\mathbb{R}^{3}$ as a
space-time model, showed that the relativistic quantum field theory inherently
forbids faster-than-light communication, \cite{Eberhard et al}. This result is
known as the no-communication theorem. It preserves the principle of causality
in quantum mechanics and ensures that information transfer does not violate
special relativity by exceeding the speed of light. It is relevant to mention
that the no-communication theorem does not rule out the possible superluminal
speed in $\mathbb{R}\times\mathbb{Q}_{p}^{3}$. Indeed, the study of the
possible superluminal speed under the hypothesis of the space-time model
$\mathbb{R}\times\left(  \mathbb{R}\times\mathbb{Q}_{p}\right)  ^{3}$ is an
open problem.

As we already pointed out in \cite{Zuniga-QM-2}, the assumption that space
discreteness implies the violation of Einstein's causality does not
immediately imply the possibility that one observer can transmit information
to another observer exceeding the speed of light. The construction of a device
that exploits the violation of the Einstein causality is an entirely different
problem from the existence of the violation phenomena.

Finally, we mention that in adelic QM, the configuration space is the product
$\mathbb{R}\times\mathbb{A}_{\mathbb{Q}}$, where $\mathbb{A}_{\mathbb{Q}}$ is
the ring of finite adeles, which consists of sequences $\left\{
x_{p}\right\}  _{p\geq2}$ from $%
{\textstyle\prod\nolimits_{p\geq2}}
\mathbb{Q}_{p}$ such that $x_{p}\in\mathbb{Z}_{p}$ for $p$ sufficiently large.
Dragovich and his collaborators have studied several quantum models on the
infinite-dimensional space $\mathbb{R}\times\mathbb{A}_{\mathbb{Q}}$; see,
e.g., \cite{AQM-1}-\cite{AQM-3}. Extra dimensions in physical models represent
a nontrivial feature. Our solution to the measurement problem relies on extra
dimensions. As we already noted, the $p$-adic QM describes continuous-time
quantum walks, including those that stem from graphs. The physical
implementation of these random walks generally involves particles or
excitations on networks of size scales ranging from a few nanometers to
micrometers. Therefore, it naturally leads the author to ask whether extra
dimensions exist at the nanoscale. This fact matches Vafa's proposal for the
existence of at least one extra spatial dimension at the nanoscale,
\cite{Vafa}.\bigskip

\textbf{Data availability statement}

No new data were created or analyzed in this study. Data sharing is not
applicable to this paper.

\end{document}